\documentclass[aip,amsmath,amssymb,reprint,groupedaddress,floatfix]{revtex4-1}
\usepackage[utf8]{inputenc}
\usepackage[T1]{fontenc}
\usepackage{graphicx}
\usepackage{amssymb}
\usepackage{nicefrac}
\usepackage{xcolor}
\usepackage{here}
\usepackage{tabularx}
\usepackage{bbm}

\usepackage{times}
\usepackage[colorinlistoftodos,prependcaption,textsize=small]{todonotes}

\usepackage{amsmath}

\usepackage{hyperref}

\hypersetup{
  colorlinks=true,
citecolor=black,
linkcolor=black,
urlcolor=black
  }

\newcommand{\bdt}[1]{\textcolor{black}{#1}}

\newcommand{\mfpt}[0]{\mathcal{T}}

\begin{document}
\preprint{AIP/123-QED}

\title{Inertial L\'evy flights in bounded domains}

\author{Karol Capa{\l}a}
\email{karol@th.if.uj.edu.pl} \affiliation{Institute of Theoretical Physics, and Mark Kac Center for Complex Systems
Research, Jagiellonian University, ul. St. {\L}ojasiewicza 11,
30--348 Krak\'ow, Poland}

\author{Bart{\l}omiej Dybiec}
\email{bartek@th.if.uj.edu.pl} \affiliation{Institute of Theoretical Physics, and Mark Kac Center for Complex Systems
Research, Jagiellonian University, ul. St. {\L}ojasiewicza 11,
30--348 Krak\'ow, Poland}

\date{\today}

\begin{abstract}
The escape from a given domain is one of the fundamental problems in statistical physics and the theory of stochastic processes.
Here, we explore properties of the escape of an inertial particle driven by L\'evy noise from a bounded domain, restricted by two absorbing boundaries.
Presence of two absorbing boundaries assures that the escape process can be characterized by the finite mean first passage time.
The detailed analysis of escape kinetics shows that properties of the mean first passage time for the integrated Ornstein--Uhlenbeck process driven by L\'evy noise are closely related to properties of the integrated L\'evy motions which, in turn, are close to properties of the integrated Wiener process.
The extensive studies of the mean first passage time were complemented by examination of the escape velocity and energy  along with their sensitivity to initial conditions.
\end{abstract}

\pacs{
 05.40.Fb, 
 05.10.Gg, 
 02.50.-r, 
 02.50.Ey, 
 }

\maketitle

\textbf{
Widely studied, first escape and first arrival processes form the basis of multiple physical phenomena with practical applications.
Extensive exploration of the first escape of an inertial particle from bounded domains under the action of symmetric L\'evy noises reveals universality, as measured by the mean first passage time, of escape kinetics in equilibrium and non-equilibrium domains.
These similarities are due to the continuity of the integrated processes, which partially reduce the significance of the discontinuity of velocity in underdamped systems driven by L\'evy noises.
However, the study of other escape characteristics like velocity or energy at the moment of first escape shows their high sensitivity to the stability index $\alpha$ as well as a dependence on initial conditions.
These studies demonstrate potentially counterintuitive properties of escape kinetics as, for instance, the slowest escape does not need to correspond to the lowest median of the escape energy.
}

\section{Introduction}

Stochastic methods \cite{horsthemke1984,gardiner2009} are widely used in the description of various systems which are too complicated to treat them exactly, or their dynamics contain an intrinsic random component.
The Newton equation plays a fundamental role in classical mechanics.
It describes the fully deterministic dynamics, nevertheless it can be easily extended to account for random perturbations.
The Newton equation supplemented by the random component is referred to as the Langevin equation \cite{coffey2012langevin}.
Typically, it is assumed that the noise is white and Gaussian, but multiple non-white or non-Gaussian extensions have been suggested \cite{klafter1987,klafter1996}.

The Langevin equation \cite{coffey2012langevin} is studied in two main regimes: the overdamped regime and in the regime of full (underdamped) dynamics.
In the overdamped domain, a random walker is fully characterized by its position only, while in the underdamped regime by position and velocity.
Therefore, overdamped situations are simpler to analyze than the full dynamics.
Nevertheless, already the overdamped Langevin equation can be used to describe and explain various noise-induced effects, like noise-enhanced stability \cite{agudov2001,dubkov2004,valenti2015}, resonant activation \cite{devoret1984,doering1992} and stochastic resetting \cite{evans2011diffusion,evans2011diffusion-jpa,evans2020stochastic}.
These phenomena do not exhaust all noise induced effects \cite{astumian1998,reimann2002,gammaitoni2009}, but they are the most relevant in the context of current research, where we focus on the problems of first escape from bounded domains \cite{benichou2005a}.

In the overdamped regime, under the action of the Gaussian white noise, one can study the free motion, which corresponds to the Wiener process \cite{gardiner2009}, or motion in a force field, e.g., Ornstein--Uhlenbeck process \cite{horsthemke1984}.
Escape properties of such random motions in bounded and half-bounded domains are well known and widely studied \cite{borodin2002} also in non-equilibrium realms \cite{palyulin2019first}.
Contrary to the overdamped motion,
on the one hand, the regime of full dynamics is more intuitive, as it incorporates velocity and position.
Consequently, some of the everyday intuitions can be easily transferred to provide a qualitative understanding of stochastic dynamics.
On the other hand, such processes are more complex to analyze and simulate.
Regime of underdamped (full) dynamics can be also referred to as the integrated process \cite{barndorff2003integrated} or randomly accelerated process \cite{bicout2000absorption,theodore2014first} since typically it is assumed that the noise affects the velocity evolution only while the position is the integral of the velocity.
Moreover, in the full regime it is possible to study undamped \cite{goldman1971first,bicout2000absorption,masoliver1996exact,masoliver1995exact,lefebvre1989first,hesse2005first} and damped motions \cite{hintze2014small,hesse1991one,magdziarz2008fractional}.
Randomly accelerated motion of a free particle corresponds to the integrated Wiener process, while the damped motion to the integrated Ornstein--Uhlenbeck process.
Action of the additional deterministic force results in the forced motions  \cite{srokowski2012anomalous,bai2017escape,capala2020underdamped}.

Escape processes from bounded (interval), and semibounded (half-line) domains are very different.
In the regime of Markovian diffusion, the escape from a finite interval is characterized by the finite mean first passage time and an exponential distribution of first passage times \cite{dybiec2017levy}.
At the same time, the escape from a half-line cannot be characterized by the mean first passage time as it diverges.
The first passage time density is given by the L\'evy--Smirnoff (inverse-Gaussian) distribution \cite{redner2001,samorodnitsky1994,chechkin2003b,koren2007,koren2007b} and it has power-law asymptotics with the exponent $-3/2$.
The asymptotics of the first passage time density is universal as the same tail asymptotics is recorded for any symmetric Markovian drivings \cite{sparre1954,sparre1953,chechkin2003b,dybiec2016jpa}.
The similar generality is observed for the integrated Ornstein--Uhlenbeck process driven by weak L\'evy noise \cite{hintze2014small}.
In the overdamped regime, exit conditions are imposed on the position, as it is only one possibility.
In the underdamped regime there are more options, since stopping conditions can be imposed on the position \cite{lefebvre1989moment,duck1986first,lefebvre1989first} or on the velocity \cite{hesse1991one,hintze2014small,hesse2005first}.

Here, we study the properties of full (underdamped) dynamics in bounded domains restricted by two absorbing boundaries with the absorption condition imposed on the position.
We relax the assumption regarding the noise type, therefore the driving noise can be of the more general $\alpha$-stable type, containing the white Gaussian noise as a special case \cite{janicki1996,samorodnitsky1994,dubkov2008}.
Consequently, our studies extend the works on the integrated Ornstein--Uhlenbeck process driven by L\'evy noise \cite{hintze2014small} to domains restricted by two absorbing boundaries.
\bdt{
Our studies rely on numerical methods, because, to the best of our knowledge, the analytical solution for the MFPT for the integrated Ornstein--Uhlenbeck process driven by the L\'evy noise is unknown.
Contrary to the underdamped case, in the regime of the overdamepd dynamics analytical results are known not only for symmetric drivings \cite{getoor1961}  but also for the escape under action of asymmetric L\'evy flights \cite{padash2019first,padash2020first}.
}
The model under study is presented in the next section (Sec.~\ref{sec:model} -- Model).
Results of computer simulations are provided in Sec.~\ref{sec:results} (Results).
The paper is closed with Summary and Conclusions  (Sec.~\ref{sec:summary}).
Technical information is moved to the Appendices.

\section{Model \label{sec:model}}

We study the archetypal, underdamped L\'evy noise-driven escape of a free particle from the finite interval $[-l,l]$.
The Langevin equation \cite{uhlenbeck1930theory,risken1996fokker} describing the motion of a single particle reads
\begin{equation}
m\ddot{x}(t)= - \gamma \dot{x}(t) + \sigma \zeta (t),
\label{eq:full-langevin}
\end{equation}
where $x$ is the particle position ($x \in [-l,l]$).
$\zeta(t)$ stands for the symmetric $\alpha$-stable (L\'evy type) noise, and $\sigma$ measures the strength of fluctuations.
The L\'evy noise is the formal time derivative of the $\alpha$-stable motion $L(t)$, see Ref. \onlinecite{janicki1994b}, with the characteristic function given by
\begin{equation}
 \phi(k)=\langle \exp[i k L(t)] \rangle=\exp\left[ - t |k|^\alpha \right].
 \label{eq:fcharakt}
\end{equation}
In the Eq~(\ref{eq:fcharakt}), $\alpha$ ($0 < \alpha \leqslant 2$) stands for the stability index, which controls the tail asymptotics of $\alpha$-stable densities \cite{samorodnitsky1994,janicki1996}. For $\alpha<2$, asymptotic behavior is of the power-law type, with the exponent $-(\alpha+1)$.
The case of $\alpha=2$ corresponds to the Gaussian white noise, i.e., $\langle \zeta(t) \zeta(s) \rangle_{\alpha=2}= \delta(t-s)$.
Furthermore, using the transformation
\begin{equation}
\left\{
\begin{array}{lcl}
\tilde{x} & = &  x/l, \\
\tilde{t} & = & \gamma t/m
\end{array}
\right.,
\end{equation}
Eq.~(\ref{eq:full-langevin}) can be transformed to the dimensionless variables $\tilde{x}$ and $\tilde{t}$.
In such variables (after dropping tildes)
\begin{equation}
\ddot{x}(t)= - \dot{x}(t) + \sigma \zeta (t),
\label{eq:full-langevin-dimensionless}
\end{equation}
where  the dimensionless fluctuation strength is expressed by dimensional variables as
$$\frac{\sigma m^{1/\alpha}}{l\gamma^{(1+\alpha)/\alpha}},$$
see Appendix~\ref{sec:units}.
Consequently, in addition to the stability index $\alpha$, the only one parameter in Eq.~(\ref{eq:full-langevin-dimensionless}) is the dimensionless strength of fluctuations $\sigma$.
The case of $\gamma=0$, see Eq.~(\ref{eq:full-langevin}), should be treated separately.
In the dimensionless units, for the undamped motion ($\gamma=0$) one gets
\begin{equation}
\ddot{x}(t)=  \zeta (t),
\end{equation}
where
\begin{equation}
\left\{
\begin{array}{lcl}
\tilde{x} & = &  x/l, \\
\tilde{t} & = & t/\left[  \frac{m l}{\sigma} \right]^{\frac{\alpha}{1+\alpha}}
\end{array}
\right..
\end{equation}
Consequently, there are no free parameters in the undamped system, see Appendix~\ref{sec:units}.

In dimensionless units, the escape from the $(-l,l)$ interval is transformed into the problem of escape from the $(-1,1)$ interval.
The problem of escape is studied in the regime of the full dynamics under the action of linear friction, therefore the particle is characterized by the position $x$ and velocity $v=\dot{x}$.
The central quantity of interest is the mean first passage time (MFPT) $\mfpt(x_0,v_0)$
\bdt{
\begin{eqnarray}
\mathcal{T}(x_0,v_0) & = & \langle t_{\mathrm{fp}}(x_0,v_0) \rangle \\
&=&\langle\min\{t>0: x(0) = x_0 \wedge v(0)=v_0 \wedge |x(t)| \geqslant 1\}\rangle. \nonumber
\label{eq:mfpt-def}
\end{eqnarray}
The mean first passage time $\mfpt(x_0,v_0)$ is the average of first passage times $t_{\mathrm{fp}}(x_0,v_0)$.
The first passage time $t_{\mathrm{fp}}(x_0,v_0)$ is recorded when a particle leaves the domain of motion, i.e., the $(-1,1)$ interval, for the first time.
Since the motion is underdamped, the first passage time depends on the full initial condition $(x_0,v_0)$.
In the dimensionless units, the motion starts in the $(-1,1)$ interval, i.e., $x_0 \in (-1,1)$, while the velocity can attains any value, i.e., $v(0) \in \mathbb{R}$.}
The studied model extends examination of the exit time properties of the inertial equilibrium process driven by Gaussian white noise \cite{porra1994mean,masoliver1995exact,masoliver1996exact} to the non-equilibrium domain, i.e., to the situation where the driving noise is of the out-of-equilibrium type.

Eq.~(\ref{eq:full-langevin-dimensionless}) can be rewritten as a set of two first-order equations
\begin{equation}
    \left\{
\begin{array}{lcl}
\dot{v}(t) & = & - v(t)  + \sigma \zeta(t) \\
    \dot{x}(t) & = & v(t) \\
\end{array}
    \right..
    \label{eq:set}
\end{equation}
The first line of Eq.~(\ref{eq:set}) describes the evolution of the velocity.
The velocity equation is the analogue of the overdamped Langevin equation describing the noise-driven motion in the parabolic ($V(x)=x^2/2$)  potential \cite{chechkin2002,chechkin2003,dybiec2007d}.
Using this analogy, the velocity can attain the stationary distribution given by the $\alpha$-stable density with the same stability index $\alpha$ \cite{chechkin2002,chechkin2003,dybiec2007d} as the noise $\zeta(t)$.
The characteristic function of stationary velocity distribution reads
\begin{equation}
    \phi_v(k) = \exp\left[ -\frac{\sigma^\alpha }{\alpha}   |k|^\alpha \right],
\end{equation}
which is the characteristic function of the symmetric $\alpha$-stable distribution, see Eq.~(\ref{eq:fcharakt}), with the scale parameter $\sigma'$
\begin{equation}
\sigma' = \sigma  \alpha^{-1/\alpha}    .
\label{eq:modsigma}
\end{equation}
The asymptotic behavior of $p(v)$ is given by
\begin{eqnarray}
p(v) \sim  \sigma^\alpha  \frac{\Gamma(\alpha+1)}{\pi} \sin \frac{\pi\alpha}{2} \times \frac{1}{|v|^{\alpha+1}}.
\label{eq:asymptoti}
\end{eqnarray}
The exact shape of the velocity distribution is sensitive to the initial velocity, furthermore it can be affected by the absorption at $x=\pm 1$.
The initial velocity shifts the modal value to nonzero locations, while the absorption can efficiently inhibit the achievement of a stationary velocity distribution, making it narrower.
Nevertheless, the velocity distribution is of the $\alpha$-stable type with the same value of the stability index $\alpha$ as the driving noise, because the instantaneous velocity is a linear transformation of $\alpha$-stable variables.
Moreover, the scale parameter characterizing the instantaneous velocity distribution cannot be larger than the scale parameter characterizing the stationary velocity distribution, i.e., $\sigma\alpha^{-1/\alpha}$.

The first line of Eq.~(\ref{eq:set}) shows that $v$ is the the $\alpha$-stable analog of the Ornstein--Uhlenbeck process \cite{vankampen1981}, i.e., the so-called L\'evy-driven Ornstein--Uhlenbeck process \cite{garbaczewski2000,eliazar2005levy}.
Moreover, due to the condition $x(t)=\int^t v(s)ds$, the $x(t)$ is the so-called integrated process, i.e., the integrated L\'evy-driven Ornstein--Uhlenbeck process \cite{hintze2014small}.
In the case of $\gamma=0$, see Eq.~(\ref{eq:full-langevin}), $v(t)$ is given by the $\alpha$-stable process, while $x(t)$ is the integrated $\alpha$-stable motion.
The special case of $\alpha=2$ corresponds to: the integrated Ornstein--Uhlenbeck process ($\gamma>0$) or the integrated Wiener process ($\gamma=0$).
Analogously, $\alpha=1$ gives rise to the integrated Cauchy process ($\gamma=0$) or the integrated Ornstein--Uhlenbeck--Cauchy process ($\gamma>0$).
The stopping condition, see Eq.~(\ref{eq:mfpt-def}), is imposed on the particle position.
The system described by Eq.~(\ref{eq:full-langevin-dimensionless}), is studied as long as $|x| < 1$.


\section{Results\label{sec:results}}

The model described by Eq.~(\ref{eq:full-langevin-dimensionless}) is studied by means of computer simulations.
The velocity part, containing the $\alpha$-stable noise, is approximated using the Euler-Maruyama scheme \cite{janicki1994,janicki1996}, while the spatial part is constructed trajectory-wise~\cite{risken1996fokker}.
The MFPTs $\mfpt(x_0,v_0)$ are calculated as the average value of the \bdt{collected} first passage times $t_{\mathrm{fp}}(x_0,v_0)$.
\bdt{
Each first passage time $t_{\mathrm{fp}}(x_0,v_0)$ is estimated from a single  trajectory $x(t)$ ($x(0)=x_0$ and $v(0)=v_0$), which is simulated as long as $|x(t)| < 1$.
The averaging is performed over the ensemble of $N=10^6$ first passage times obtained from $N$ trajectories constructed with the integration time step $\Delta t=10^{-3}$.
}

\begin{figure}
    \centering
    \includegraphics[width=0.9\columnwidth]{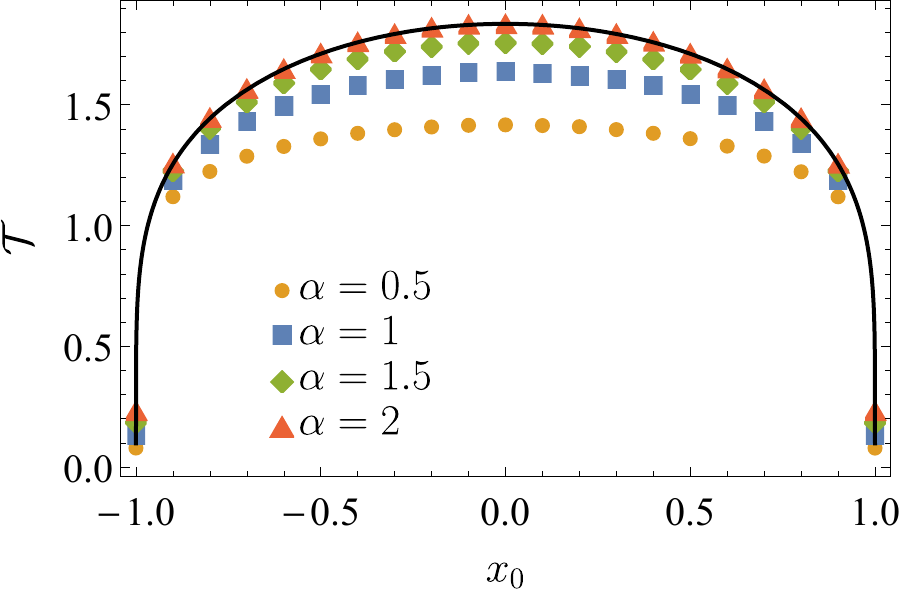}
    \caption{The mean first passage time (MFPT) $\mathcal{T}(x_0,0)$ for integrated $\alpha$-stable motions. Various points correspond to various values of the stability index $\alpha$ ($\alpha\in\{0.5,1,1.5,2\}$) from the lowest to highest MFPT respectively.
    Solid lines show the theoretical formula for $\alpha=2$ (integrated Wiener process), see Eq.~(\ref{eq:integratedwiener-mfpt-dm}).
    }
    \label{fig:integratedlevymotion}
\end{figure}

The main quantity characterizing the escape kinetics is the mean first passage, see Eq.~(\ref{eq:mfpt-def}).
The mean first passage time depends on both the value of the stability index $\alpha$ and the strength of fluctuations $\sigma$.
Since the motion is restricted to the $(-1,1)$ interval, the initial position $x_0$ belongs to $(-1,1)$.
There are no constraints on the initial velocity $v_0$.
It can be directed towards any of the absorbing boundaries.

We start our analysis with the undamped motion, i.e., with the integrated $\alpha$-stable motion (Sec.~\ref{sec:iw}), which for $\alpha=2$ corresponds to the integrated Wiener process.
Next, we switch to the problem of inertial damped motion, i.e., the integrated Ornstein--Uhlenbeck process driven by L\'evy noise (Sec.~\ref{sec:iou}).

\subsection{Integrated $\alpha$-stable motion\label{sec:iw}}

The integrated $\alpha$-stable motion corresponds to $\gamma=0$ in Eq.~(\ref{eq:full-langevin}).
In dimensionless units, see Appendix~\ref{sec:units}, it is described by the following Langevin equation
\begin{eqnarray}
\ddot{x}(t)= \zeta(t).
\label{eq:integratedalphasrable}
\end{eqnarray}
Eq.~(\ref{eq:integratedalphasrable}) with $\alpha=2$ describes the integrated Wiener process.
More precisely, $\zeta_{\alpha=2}(t)=\sqrt{2}\xi(t)$ (where $\xi(t)$ stands for the standard Gaussian white noise) as the $\alpha$-stable density with $\alpha=2$ is the normal distribution with the standard deviation $\sqrt{2}$.
Such a process has been studied in Refs.~\onlinecite{masoliver1995exact,masoliver1996exact}, where the exact, up to quadrature, formula for the MFPT with any allowed value of $x_0$ and $v_0$ has been derived.
The general formula \cite{masoliver1995exact,masoliver1996exact} significantly simplifies for $v_0=0$, see Eq.~(\ref{eq:integratedwiener-mfpt}).
After transformation of the $[0,l]$ setup \cite{masoliver1995exact,masoliver1996exact} to the $[-l,l]$ and passing to dimensionless variables, see Appendix~\ref{sec:units}, the formula for the MFPT with $v_0=0$ reads
\begin{eqnarray}
\label{eq:integratedwiener-mfpt-dm}
\mathcal{T}(x,0) & = &
\frac{2}{3^{1/6} \Gamma(7/3)}
\left[ \frac{1+x}{2}  \right]^{1/6}
\left[ \frac{1-x}{2}  \right]^{1/6} \\
& \times &
\left\{
{}_2F\left(  1,-\frac{1}{3};\frac{7}{6}; \frac{1+x}{2}  \right)
+
{}_2F\left(  1,-\frac{1}{3};\frac{7}{6}; \frac{1-x}{2}  \right)
\right\} \nonumber,
\end{eqnarray}
where ${}_2F\left(  a,b ; c; x  \right)$ is the Gauss hypergeometric function \cite{abramowitz1964handbook}.
For more details see Appendix~\ref{sec:ibm}.

Fig.~\ref{fig:integratedlevymotion} shows dependence of the mean first passage time on $x_0$ for various values of the stability index $\alpha$ ($\alpha\in\{0,5,1,1.5,2\}$ -- from bottom to the top) with the fixed initial velocity $v_0$, i..e, $v_0=0$.
The solid line depicts the MFPT  for $\alpha=2$ given by Eq.~(\ref{eq:integratedwiener-mfpt-dm}).
With the decreasing value of the stability index $\alpha$ (in dimensionless variables) the MFPT decreases, i.e., the escape on average becomes faster.
Nevertheless, the qualitative dependence of the MFPT on $x_0$ with various values of $\alpha$ is similar.
In dimensional units, the order of MFPT curves is sensitive to the system parameters.

\subsection{Integrated Ornstein--Uhlenbeck L\'evy-driven process\label{sec:iou}}

\begin{figure}[!h]
    \centering
    \includegraphics[width=0.9\columnwidth]{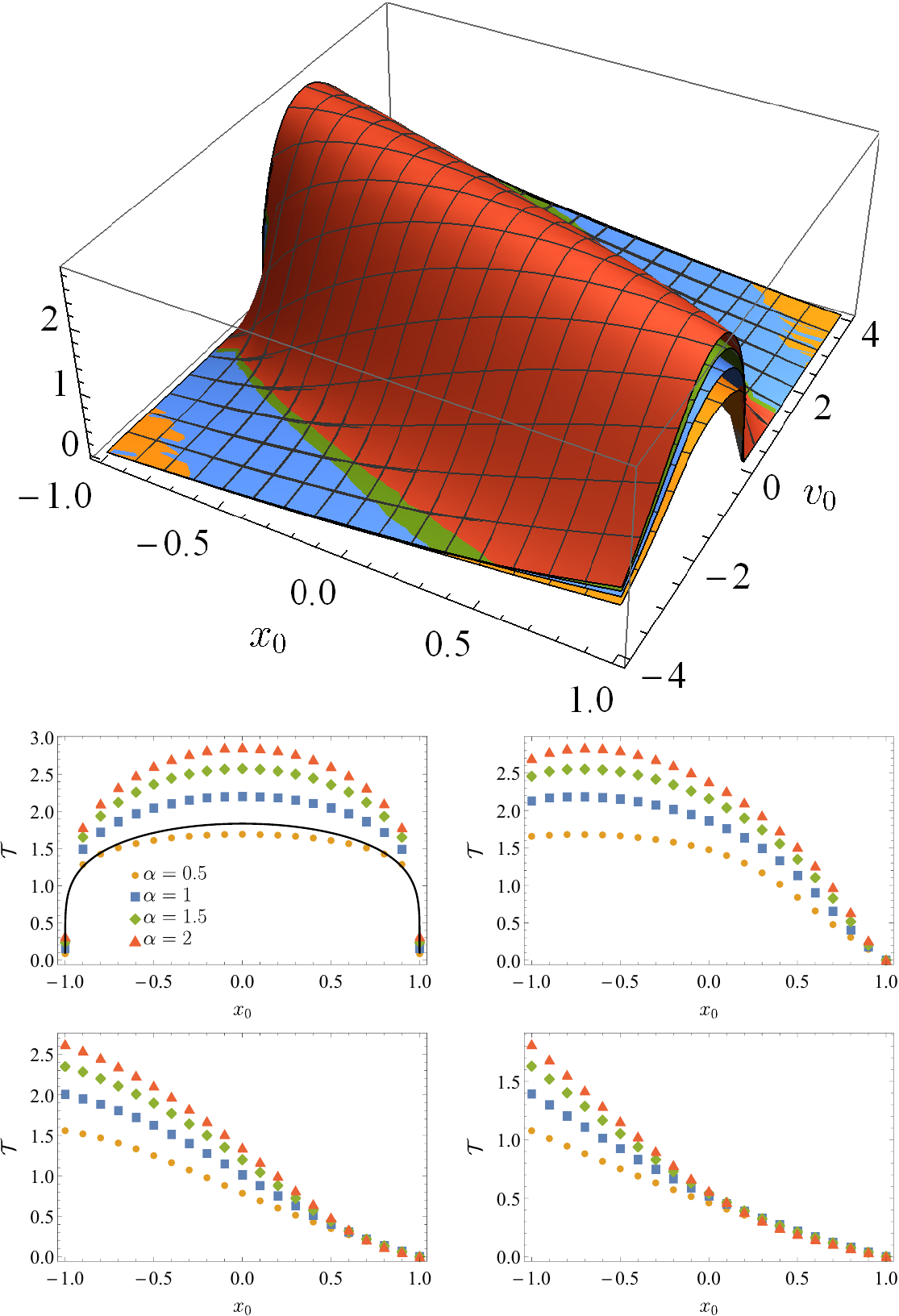}
    \caption{The MFPT $\mathcal{T}(x_0,v_0)$ as a function of the initial condition $(x_0,v_0)$ for various values of the stability index $\alpha$ ($\alpha\in\{0.5,1,1.5,2\}$) from the lowest to highest MFPT respectively.
    Bottom part shows cross-sections for $v_0=0$ (top left), $v_0=1$ (top right), $v_0=2$ (bottom left) and $v_0=3$ (bottom right).
    The scale parameter $\sigma$ is set to $\sigma=1$.
    }
    \label{fig:MFPT}
\end{figure}

\begin{figure}[!h]
    \centering
    \includegraphics[width=0.9\columnwidth]{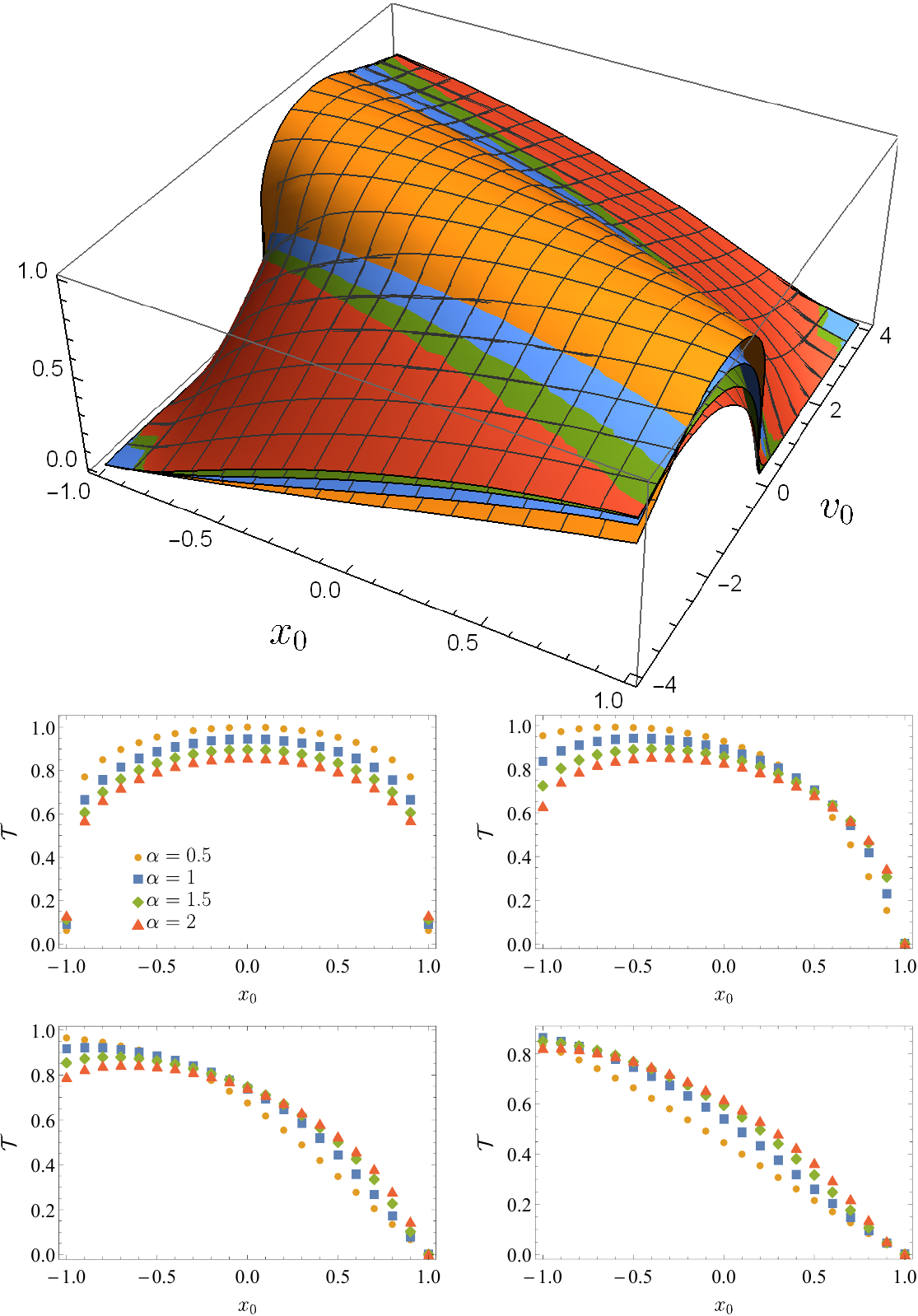}
    \caption{The same as in Fig.~\ref{fig:MFPT} for $\sigma=4$.
    }
    \label{fig:MFPTSigma4}
\end{figure}

Fig.~\ref{fig:MFPT} depicts the MFPT for the damped motion described by Eq.~(\ref{eq:full-langevin-dimensionless}) as the function of the initial condition ($x_0,v_0$) for \bdt{representative} values of the stability index $\alpha$ ($\alpha\in\{0.5,1,1.5,2\}$ from bottom to top).
Moreover, the bottom panel of Fig.~\ref{fig:MFPT} shows cross-sections corresponding to various initial velocities: $v_0=0$ (top left), $v_0=1$ (top right), $v_0=2$ (bottom left) and $v_0=3$ (bottom right).
Additionally the cross-section for $v_0=0$ is accompanied by the exact curve showing MFPT for the integrated Wiener process, see Eq.~(\ref{eq:integratedwiener-mfpt-dm}).
Analogously, like for the integrated $\alpha$-stable motion, the MFPT surfaces are symmetric with respect to exchange $(x_0,v_0)$ with $(-x_0,-v_0)$, i.e.,
\begin{equation}
\mfpt(x_0,v_0)=\mfpt(-x_0,-v_0).
\label{eq:mfpt-symmetry}
\end{equation}
The relation given by Eq.~(\ref{eq:mfpt-symmetry}) arises due to symmetry of the experimental setup and system dynamics.
From Fig.~\ref{fig:MFPT}, especially from cross-sections, it is clearly visible that the MFPT is sensitive to the exact value of the stability index $\alpha$.
For $\sigma=1$, decrease in the value of the stability index $\alpha$ facilitates the escape kinetics.
For $v_0=0$ MFPT curves are symmetric along $x_0=0$, moreover quantitative dependence of MFPT on $x_0$ with $v_0=0$ is the same as for the integrated Wiener process with $v_0=0$.
The initial (positive) velocity significantly accelerates the escape process for positive $x_0$ and slows down escapes with negative $x_0$.

\begin{figure}[!h]
    \centering
    \includegraphics[width=0.9\columnwidth]{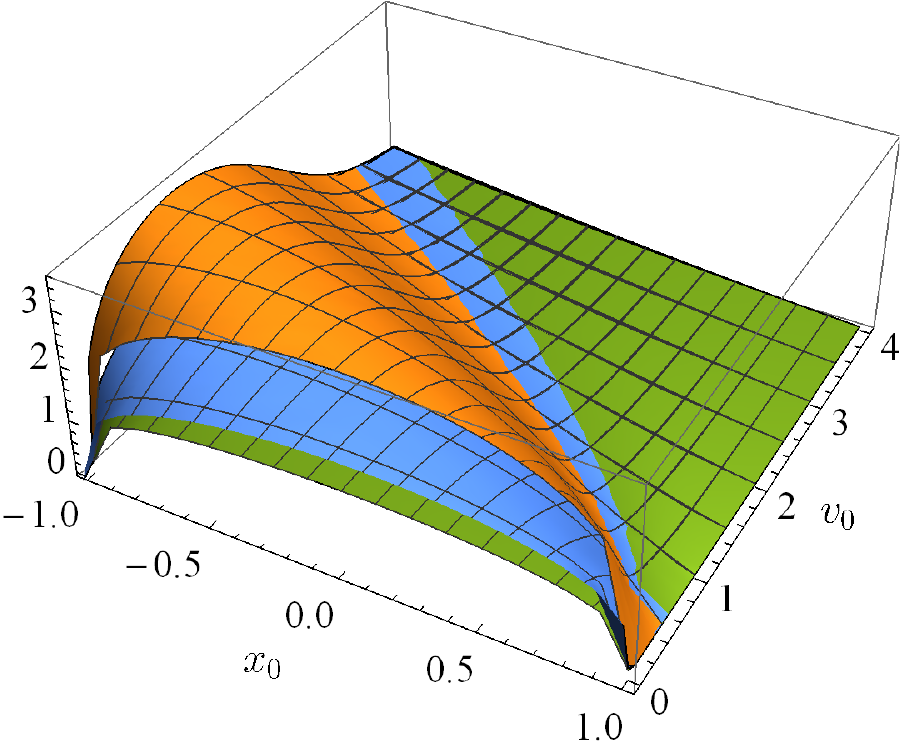}
    \caption{The MFPT as a function of the initial condition $(x_0,v_0)$ for the fixed value of the stability index $\alpha=1$ (Cauchy noise) and various strengths of fluctuations $\sigma$  ($\sigma\in\{0.5,1,2\}$ (orange, blue, green).}
    \label{fig:MFPTSigma}
\end{figure}

Fig.~\ref{fig:MFPTSigma4} presents dependence of the MFPT on the initial conditions for the noise strength $\sigma=4$, which is significantly larger than the one considered in Fig.~\ref{fig:MFPT}.
The change in $\sigma$ not only facilitates escape in comparison to $\sigma=1$, but changes the order of surfaces in the 3D plot and curves in cross-sections.
For $\sigma=4$, with $v_0=0$, the fastest escape is recorded for Gaussian noise ($\alpha=2$).
Moreover, contrary to smaller $\sigma$ (e.g., $\sigma=1$), this time MFPT is decreasing function of $\alpha$.
Subsequent Fig.~\ref{fig:MFPTSigma} demonstrates that for the fixed value of the stability index $\alpha$, alterations in the scale parameter can produce well visible changes in the mean first passage time, especially for not too large initial velocities.

\begin{figure}[!h]
    \centering
    \includegraphics[width=0.9\columnwidth]{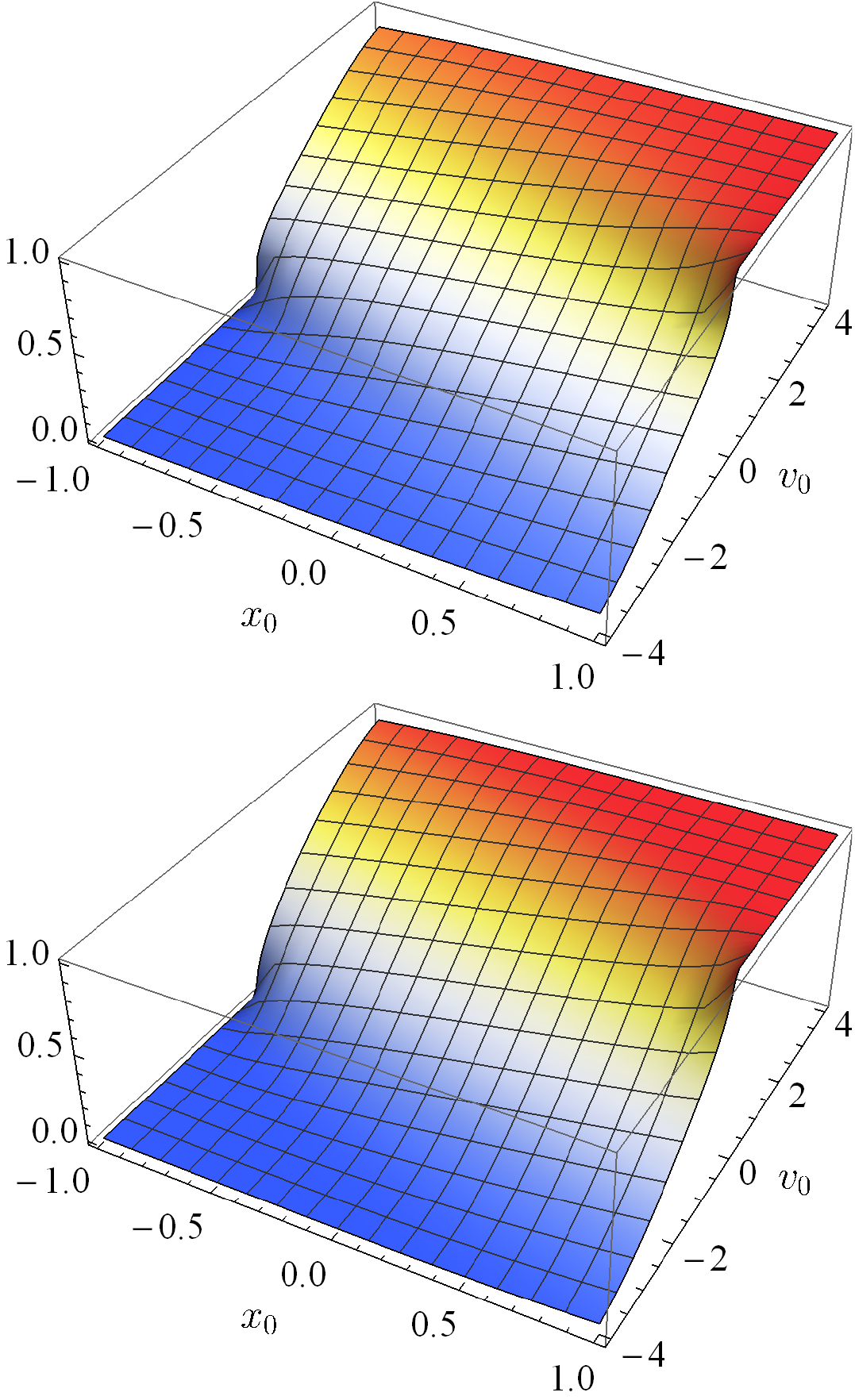}
    \caption{The probability $\pi_R$ that the particle escapes from the finite interval $(-1,1)$ through the right boundary, i.e., $x=1$, as a function of the initial condition ($x_0,v_0$) for the fixed value of the stability index  $\alpha=1$ (Cauchy noise -- top panel) and $\alpha=2$ (Gaussian noise -- bottom panel).
    The scale parameter $\sigma$ is set to $\sigma=1$.
    }
    \label{fig:split}
\end{figure}

The subsequent Fig.~\ref{fig:MFPTSigma} explores the sensitivity of the mean first passage time  to the strength of fluctuations under the action of the Cauchy noise ($\alpha$-stable noise with $\alpha=1$).
Due to the property given by Eq.~(\ref{eq:mfpt-symmetry}), Fig.~\ref{fig:MFPTSigma} shows results for $v_0>0$ only.
The highest sensitivity to the fluctuation strength is recorded, when a particle starts its motion far from the boundary which is crossed during the escape from the domain of motion, see Fig.~\ref{fig:split}.
With the increasing $|v_0|$ the level of sensitivity decreases.
Finally, for very large $v_0$ results with various scale parameters $\sigma$ are indistinguishable.

The particle escaping from the $(-1,1)$ interval can exit via the left or right boundary.
The tendency to exit via a particular boundary can be quantified by the splitting probability $\pi_R$.
For instance, $\pi_R$ measures the fraction of escapes via the right boundary.
At the same time, the fraction of escapes via the left boundary can be calculated as $\pi_L=1-\pi_R$.
The dependence of the splitting probability $\pi_R$ on the initial condition ($x_0,v_0$) is depicted in Fig.~\ref{fig:split}.
The top panel shows the results for the Cauchy ($\alpha=1$) noise, while the bottom one for the Gaussian ($\alpha=2$) noise.
Positive initial velocity favors escape via the right boundary.
This tendency is especially visible for the initial positions near the right boundary.
The change in the value of the stability index $\alpha$ from $\alpha=1$ (top panel) to $\alpha=2$ (bottom panel) does not change the qualitative dependence of the splitting probability on the initial condition.
Only small quantitative changes are visible in the situation when a motion starts near the boundary with the initial velocity pointing to the more distant boundary.

The stopping condition is imposed on the particle position $x(t)$, see Eq.~(\ref{eq:mfpt-def}), which is continuous as the integral of the velocity.
Due to the continuity of trajectories, every trajectory hits the absorbing boundary.
Therefore, the last hitting point density is given by $\pi_R \delta(x-1)+(1-\pi_R)\delta(1-x)$, where $\pi_R$ is the probability to escape through the right boundary.
Nevertheless, the escape is not only characterized by the splitting probability, see Fig.~\ref{fig:split}, but also by the exit velocity, i.e., the instantaneous velocity at the moment of the first escape.
The distribution of the exit velocities is determined by the instantaneous velocity distribution, which is given by the symmetric $\alpha$-stable density, and the initial condition ($x_0,v_0$).
The initial condition is capable of introducing asymmetry to the exit velocity distribution.
For $x_0=0$ with $v_0=0$, the distribution of exit velocities is symmetric.

\begin{figure}[!h]
    \centering
    \includegraphics[width=0.9\columnwidth]{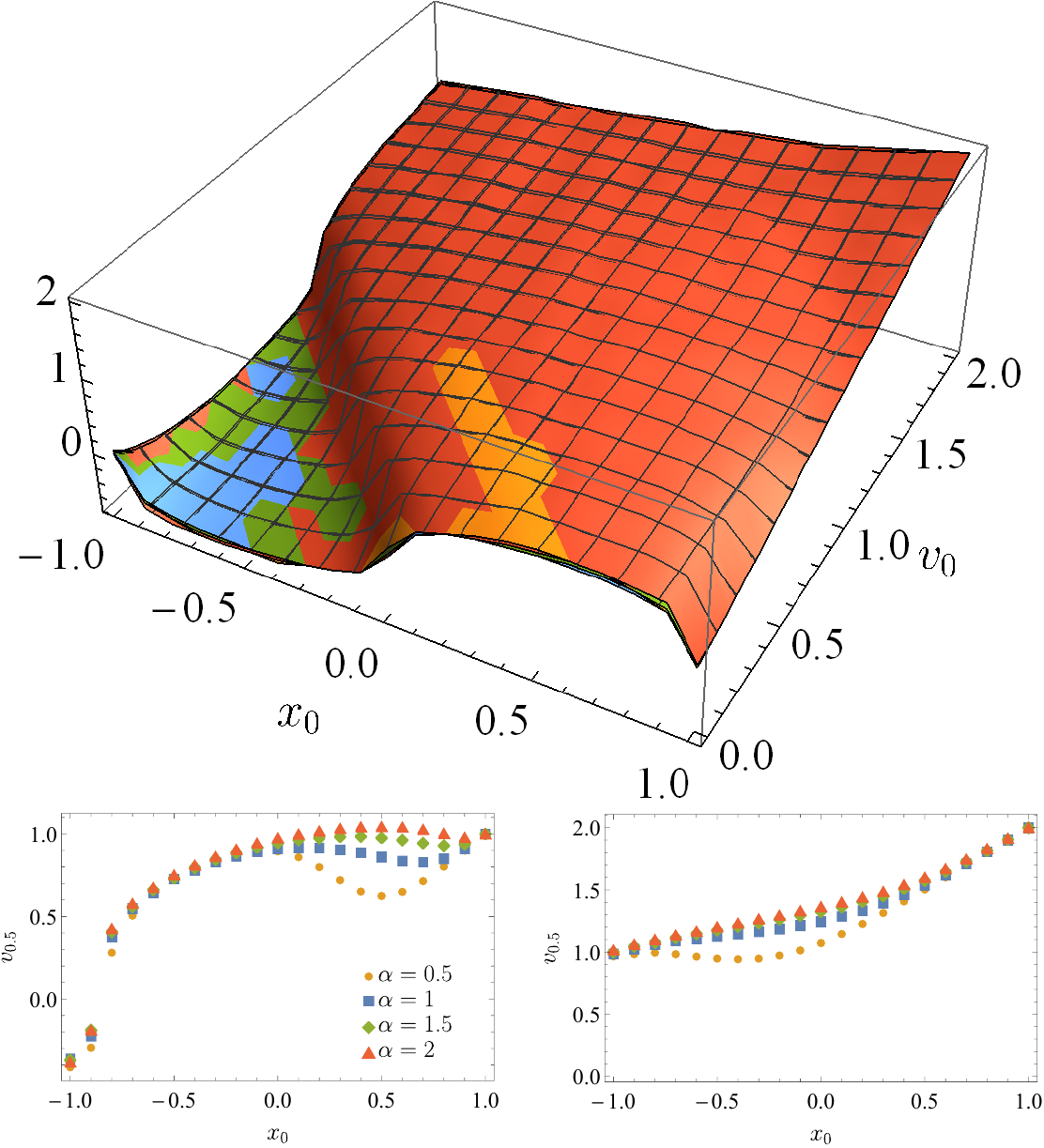}
    \caption{Medians of the exit velocity $v_{0.5}$ as a function of the initial condition ($x_0,v_0$) for  various values of the stability index $\alpha$ ($\alpha\in\{0.5,1,1.5,2\}$ (orange, blue, green, red))  (top panel) and cross-sections for $v_0=1$ (bottom left) and $v_0=2$ (bottom right).
    The scale parameter $\sigma$ is set to $\sigma=1$.
    }
    \label{fig:predkosciMediana}
\end{figure}

Fig.~\ref{fig:predkosciMediana} presents the median ($v_{0.5}$) of exit velocities as a function of the initial condition ($x_0,v_0$).
Various surfaces in the top panel of Fig.~\ref{fig:predkosciMediana} correspond to various values of the stability index $\alpha$ ($\alpha\in\{0.5,1,1.5,2\}$).
$\alpha$-stable distributions with such values of stability indices  are very different.
Nevertheless, the medians of the exit velocities are quite similar, which is further corroborated by cross-sections depicted in the bottom panel of Fig.~~\ref{fig:predkosciMediana}.
\bdt{Unexpectedly, for some values of the initial velocity, e.g., $v_0=1$, there is a local minimum of the median ($v_{0.5}$) of the exit velocity, see bottom row of Fig.~\ref{fig:predkosciMediana}.
As it is visible from the splitting probability, see Fig.~\ref{fig:split}, cross sections in Fig.~\ref{fig:predkosciMediana} correspond to the situations when the first escape takes place via both absorbing boundaries. 
The level of competition is sensitive both to the initial position and the initial velocity as they determine which part of the velocity distribution is responsible for the final jump. 
If a particle is close to the boundary it can (most likely) leave the domain of motion with a small velocity through the closest boundary or it can escape via the distant boundary with a large velocity with the opposite sign. 
The chances of escaping via the more distant boundary are decreasing with the increasing $\alpha$ making local minima of $v_{0.5}$ shallower and shifted to larger $x_0$, see bottom left panel of Fig.~\ref{fig:predkosciMediana}.
}

\begin{figure}[!h]
    \centering
    \includegraphics[width=0.9\columnwidth]{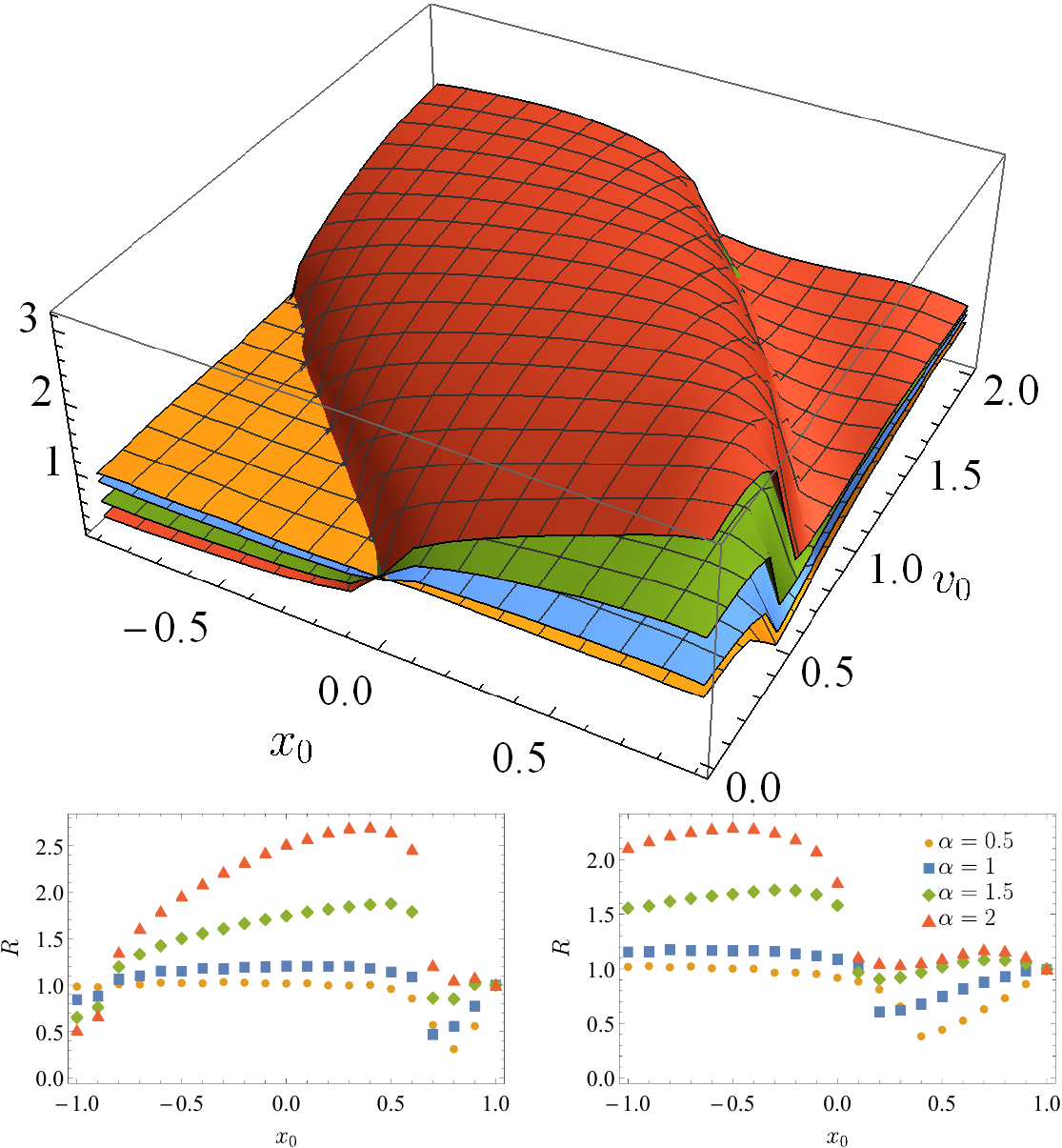}
   \caption{Interquantile widths ratios  $\mathcal{R}=(v_{0.5}-v_{0.1})/(v_{0.9}-v_{0.5})$ for $\alpha\in\{0.5,1,1.5,2\}$ (top panel) and cross-sections for $v_0=1$ (bottom left) and $v_0=2$ (bottom right).
    The scale parameter $\sigma$ is set to $\sigma=1$.
    }
    \label{fig:predkosciSymetria}
\end{figure}

\bdt{In order to further explore}
properties of exit velocity, we have calculated the ratio of interquantile widths
\begin{equation}
\mathcal{R}     = \frac{v_{0.5}-v_{0.1}}{v_{0.9}-v_{0.5}},
\label{eq:width}
\end{equation}
where $v_{\dots}$ indicates quantiles of a given order $q$ ($0<q<1$) of the exit velocity, e.g., $v_q(t)$ is defined by
\begin{equation}
 q = \int_{-\infty}^{v_q(t)} p(v;t) dv.
 \label{eq:quantile}
\end{equation}
In the above equation, $p(v;t)$ stands for exit velocity distribution.
The ratio defined by Eq.~(\ref{eq:width}) measures the fraction of widths of intervals containing $40\%$ of the exit velocities
above ($v_{0.9}-v_{0.5}$) and below ($v_{0.5}-v_{0.1}$) the median ($v_{0.5}$).
Its value reflects the symmetry of the exit velocity distribution: for $\mathcal{R}=1$, the intervals' widths in the numerator  and the denominator are the same.
If $\mathcal{R}<1$ the exit velocity density is skewed to the right, while for $\mathcal{R}>1$ to the left.
Fig.~\ref{fig:predkosciSymetria} depicts the interquantile width ratios, which show that the width ratio $\mathcal{R}$ is sensitive to the exact value of the stability index $\alpha$.
For $\alpha=0.5$ the ratio of interquantile widths is practically equal to $1$, as with decreasing $\alpha$ exit velocities become more symmetric.
Finally, in the limit of $\alpha\to 0$, the ratio $\mathcal{R}$ is equal to 1, because the escape events become position independent.
In the opposite limit of $\alpha\to 2$, the escape process strongly depends on the initial position, because it is easier to escape via the closest boundary.
Fig.~\ref{fig:predkosciMediana} and \ref{fig:predkosciSymetria} present results for $v_0>0$ because results for $v_0<0$ can be obtained by symmetry.
For instance, quantiles of order $q$ ($0<q<1$) are connected with quantiles of order $1-q$ by the relation $v_q(x_0,v_0)=-v_{1-q}(-x_0,-v_0)$ from which implies that $\mathcal{R}(-x_0,-v_0)=1/\mathcal{R}(x_0,v_0)$.

From analysis of Fig.~\ref{fig:predkosciSymetria}, \bdt{supported by the examination of the splitting probability (Fig.~\ref{fig:split}) and median of the exit velocity (Fig.~\ref{fig:predkosciMediana})}, it can be deduced that there are two mechanisms which can produce $\mathcal{R}\approx 1$.
One, the most intuitive, is related to the symmetry of the exit velocity.
\bdt{This mechanism is observed for $v_{0.5}=0$, where the instantaneous velocity distribution is symmetric and approximately half of escapes are via the left (right) boundary with  negative (positive) velocities.
Such a behavior is the  strongest for $x_0 \approx 0$ and $v_0 \approx 0$, and the decrease in $\alpha$ widens the domain where $\mathcal{R}\approx 1$.
}
\bdt{
The other mechanism is related to the initial velocity.
For large $|v_0|$ the mean first passage time is small and it is mainly determined by the initial velocity, which determines the time dependence of the median of the velocity distribution.  
The nonzero median of the velocity distribution forces the median of the position distribution to move towards one of the absorbing boundaries. 
This in turn facilitates the escape. 
Moreover, the escape time is so short that the width (as measured by the interquantile width) of the velocity distribution is small and median quite large that escapes are performed over one of the boundaries determined by the initial condition.  
In overall, the median of exit velocity is significant and, simultaneously, the exit velocities follow narrow, symmetric along the median, density making $\mathcal{R}$ again close to 1.
In other regions, where escapes are performed via both absorbing boundaries,  we can see the competition between escapes to the left and to the right. 
}

\begin{figure}[!h]
    \centering
    \includegraphics[width=0.9\columnwidth]{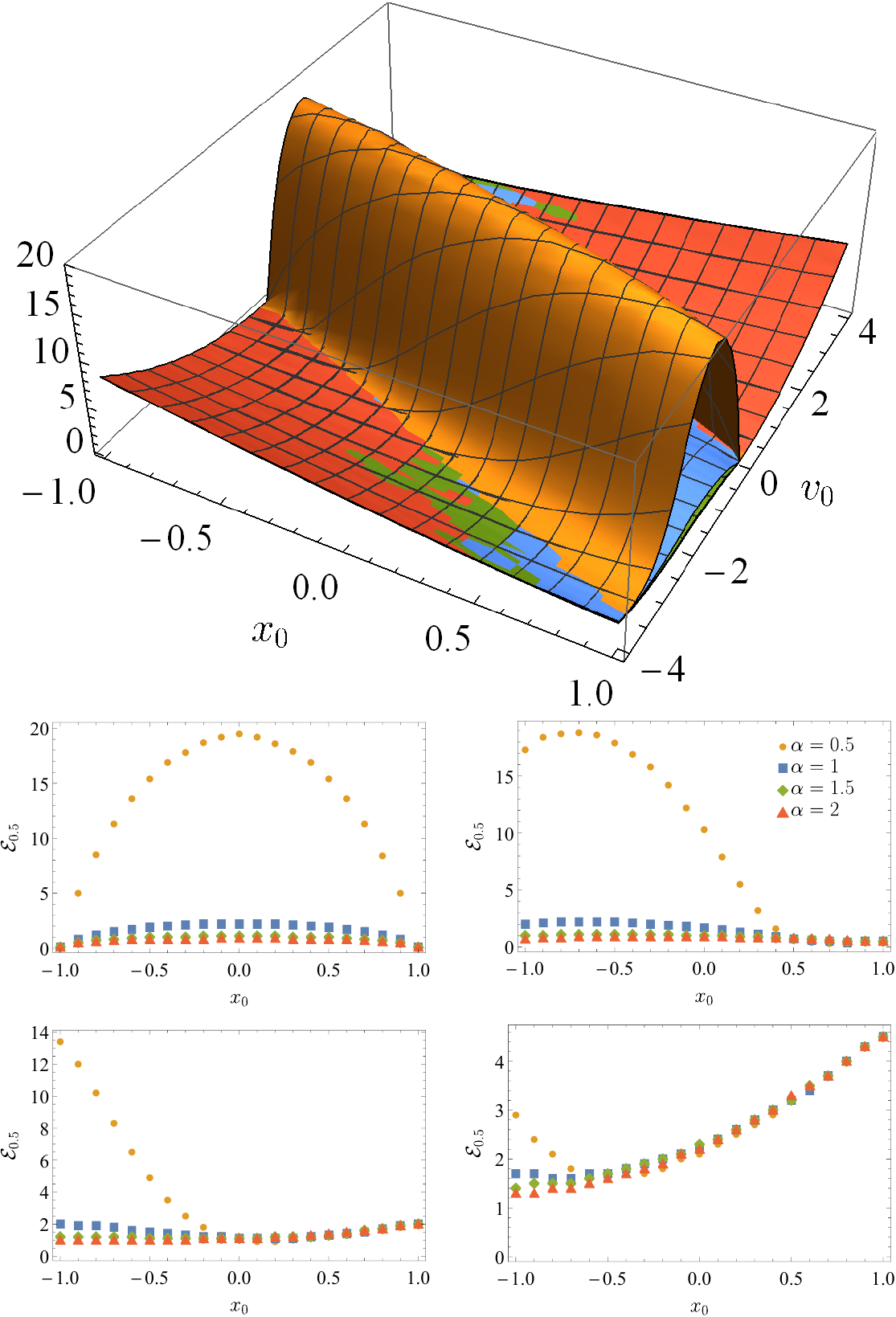}
    \caption{Median of exit energy $\mathcal{E}_{0.5}$ as a function of the initial condition ($x_0,v_0$) for various values of the stability index $\alpha$   ($\alpha \in\{0.5,1,1.5,2\}$ (orange, blue, green, red)).
    Bottom part shows cross-sections for $v_0=0$ (top left), $v_0=1$ (top right), $v_0=2$ (bottom left) and $v_0=3$ (bottom right).
    The scale parameter $\sigma$ is set to $\sigma=1$.
}
    \label{fig:energie}
\end{figure}

Finally, to further study the properties of escape kinetics, we have inspected energies at the moment of boundary crossing.
Fig.~\ref{fig:energie} shows median of the energy distributions  along with sample cross-sections.
One may think that the slowest escapes (largest MFPTs) correspond to the lowest median of the escape energy.
However, it is not the case.
For the initial conditions $(x_0,v_0)$ corresponding to longest escape time, there is a local maximum (hump) in the median of the escape energy.
Such a maximum originates due to slow escapes.
More precisely, a particle spends a lot of time within the interval.
During that time a chance for an abrupt change in the velocity increases and the particle is likely to leave the domain of motion with a large velocity (and energy).
Height of the hump in the median of energy is very sensitive to the stability index $\alpha$ and decays rapidly with its decreasing value, therefore, in the top panel of the Fig.~\ref{fig:energie}, maximum is well visible only for $\alpha=0.5$.
\bdt{
In general, for a fixed initial condition located within the hump, the median of the exit energy decays with the decreasing value of the stability index $\alpha$. 
Consequently,  the smallest $\alpha$ curve is the predominant one. 
After removing results corresponding to $\alpha=0.5$ the hump with $\alpha=1.0$ starts to prevail. 
Nevertheless, the qualitative dependence of the median of the exit energy is quite similar. 
The biggest qualitative differences start to appear when $\alpha$ approaches 2.
}

Interestingly, the well-defined hump in the median of energy distribution is placed within a gutter.
The trough rises upwards with increasing modulus of the initial velocity.
Therefore, starting from the top of the hump, with the increasing (decreasing) initial velocity median decreases, attains minimum value and finally it starts to increase.
This indicates that the increase in the value of the initial velocity $v_0$ does not always lead to the larger escape energy.
As it is clearly visible in Fig.~\ref{fig:energie}, escapes with largest initial velocities, see Fig.~\ref{fig:MFPT}, are performed with the largest energies.
Moreover, the medians of the escape energy in this case are insensitive to the stability index $\alpha$ and, therefore, they are indistinguishable in the plot.
It suggests that escapes energies for large $v_0$ are mainly controlled by the initial condition.

\section{Summary and Conclusions\label{sec:summary}}

In the weak noise limit of the integrated Ornstein--Uhlenbeck process driven by L\'evy noise, the distribution of first passage times from the half-line attains the universal form \cite{hintze2014small}, which is independent of the driving noise type.
This is related to the general properties of escape kinetics from the half-line under symmetric Markovian drivings.
Such escapes are characterized by first passage time densities with the universal power-law asymptotics predicted by the Sparre Andersen scaling \cite{sparre1954,sparre1953,chechkin2003b,dybiec2016jpa}.
Due to the heavy tail of the first passage time density, the first passage density has a power-law asymptotic with the exponent $-3/2$.
The escape from the half-line cannot be characterized by the mean first passage time as this quantity diverges.

Using numerical methods, we have studied the properties of underdamped L\'evy noise-driven escape from finite intervals restricted by two absorbing boundaries.
For such a process, contrary to the escape from the half-line, the exit time distributions have exponential asymptotics making the mean first passage time well defined characteristics.
Detailed examination of the integrated L\'evy-driven Ornstein--Uhlenbeck process indicates that the mean first passage time displays limited sensitivity to the exact value of the stability index $\alpha$.
Therefore, despite very different velocity distributions, the qualitative system properties are very close to the properties of the integrated Wiener process.
Nevertheless, the increase in the scale parameter (the only significant parameter besides $\alpha$) can differentiate results corresponding to various values of the stability index $\alpha$.
The symmetry of the domain of motion and system dynamics is responsible for additional symmetries of the mean first passage times with respect to the initial conditions.

The extensive analysis of the mean first passage time was supplemented by examination of the escape velocities and energies along with their sensitivity to the initial conditions.
On the one hand, for the L\'evy-driven Ornstein--Uhlenbeck process, medians of the escape velocity are weakly sensitive to the stability index $\alpha$.
On the other hand, analysis of the asymmetry of the escape velocity distributions show a high level of sensitivity to the stability index.
In particular, distributions change from almost always symmetric (small $\alpha$) to possibly strongly dependent on the initial condition ($\alpha \lesssim 2$).
Putting it differently, for large $\alpha$, depending on the initial condition, the escape velocity distribution can be symmetric or not.
For example, for the same initial position, the escape velocity distribution can be asymmetric (small initial velocity) and symmetric (large initial velocity).
In contrast to velocities, the studies of escape energies manifest significant sensitivity of the medians to the value of the stability index $\alpha$, especially for small initial velocities.
Finally, the median of the escape energy can decrease with the increasing value of initial velocity.

%
%
\appendix

\section{Dimensionless units\label{sec:units}}

The dimensional Langevin equation
\begin{equation}
m\ddot{x}(t)= - \gamma \dot{x}(t) + \sigma \zeta (t),
\label{eq:full-langevin-units}
\end{equation}
can be transformed to dimensionless units $\tilde{x}$ and $\tilde{t}$ by rescaling the space variable $x$ and time $t$
\begin{equation}
    \left\{
    \begin{array}{lcl}
    \tilde{x} & = & \frac{x}{x_0} \\
    \tilde{t} & = & \frac{x}{t_0} \\
    \end{array}
    \right..
\end{equation}
The noise term is a formal time derivative of the $\alpha$-stable motion, which is a $1/\alpha$ self-similar process.
Therefore, it transforms as
\begin{eqnarray}
\sigma \zeta(t) & = & \sigma \frac{dL(t)}{dt} = \sigma \frac{d}{dt} L(t_0\tilde{t})=\sigma \frac{d}{dt}t_0^{\frac{1}{\alpha}} L(\tilde{t}) \\
& = & \sigma t_0^{\frac{1}{\alpha}} \frac{d}{dt} L(\tilde{t}) = \sigma t_0^{\frac{1}{\alpha}} \frac{d\tilde{t}}{dt} \frac{d}{d\tilde{t}} L(\tilde{t}) \nonumber \nonumber \\
& = & \sigma t_0^{\frac{1}{\alpha}-1} \zeta(\tilde{t}) \nonumber.
\end{eqnarray}
At the same time standard derivatives account the following forms
\begin{equation}
    m \frac{d^2x(t)}{d^2t}= m\frac{x_0}{t_0^2} \frac{d^2\tilde{x}}{d\tilde{t}^2}
\end{equation}
and
\begin{equation}
    \gamma \frac{d x(t)}{d t}= \gamma\frac{x_0}{t_0} \frac{d\tilde{x}}{d\tilde{t}}.
\end{equation}
After substituting in Eq.~(\ref{eq:full-langevin-units}) one gets
\begin{equation}
    m\frac{x_0}{t_0^2} \frac{d^2\tilde{x}}{d\tilde{t}^2} = -\gamma\frac{x_0}{t_0} \frac{d\tilde{x}}{d\tilde{t}} + \sigma t_0^{\frac{1}{\alpha}-1} \zeta(\tilde{t}) .
\end{equation}
Dividing both sides by $m\frac{x_0}{t_0^2}$ and dropping tildes result in
\begin{equation}
     \frac{d^2 x}{d t^2} = -  \gamma\frac{t_0}{m }   \frac{d x }{dt} + \frac{ \sigma t_0^{1+\frac{1}{\alpha}} }{m x_0}  \zeta(t).
\end{equation}
Since we are interested in the exploration of escape from $[-l,l]$ interval we set $x_0$ to $l$, thus $\tilde{x}=\frac{x}{l}$.
Consequently, in the dimensionless units, we are studying escape from the $(-1,1)$ interval.
For $\gamma>0$, setting $\gamma\frac{t_0}{m }$ to $1$ one gets $t_0=\frac{m}{\gamma}$ and
\begin{equation}
    \tilde{t}=\frac{\gamma}{m} t.
\end{equation}
Moreover, for $\gamma>0$ one finds the rescaled $\tilde{\sigma}$
\begin{equation}
    \tilde{\sigma}= \frac{ \sigma t_0^{1+\frac{1}{\alpha}} }{m x_0}  = \frac{ \sigma m^{\frac{1}{\alpha}} }{l \gamma^{1+\frac{1}{\alpha}}}.
\end{equation}
The case of $\gamma=0$ needs to be considered separately.
We still use $\tilde{x}=\frac{x}{l}$, while the time transformation is found from the condition
\begin{equation}
    \frac{ \sigma t_0^{1+\frac{1}{\alpha}} }{m x_0} = 1
\end{equation}
resulting in
\begin{equation}
    t_0=  \left[  \frac{m x_0}{\sigma} \right]^{\frac{\alpha}{1+\alpha}} = \left[  \frac{m l}{\sigma} \right]^{\frac{\alpha}{1+\alpha}}.
    \label{eq:t0}
\end{equation}

In summary, in the dimensionless units, for $\gamma>0$, one has
\begin{equation}
    \ddot{x}(t)=-\dot{x}(t) + \sigma \zeta(t),
\end{equation}
while for $\gamma=0$
\begin{equation}
    \ddot{x}(t)= \zeta(t).
\end{equation}
For the sake of simplicity, in the above equations, the tildes have been dropped out.
The motion is continued as long as $|x| < 1$.
Therefore, the damped motion is characterized by the dimensionless $\sigma$ only, while in the undamped case there are no free parameters.

The $\alpha$-stable density with $\alpha=2$ reduces to the normal (Gaussian) distribution with the standard deviation $\sqrt{2}$.
Therefore, special care is required if one wants to compare results for $\alpha$-stable driving with $\alpha=2$ to results derived under action of the Gaussian white noise.
The factor $\sqrt{2}$, see Eq.~(\ref{eq:t0}), needs to be accounted for.

\section{Integrated Brownian motion (random acceleration process)\label{sec:ibm}}

In Refs.~\onlinecite{masoliver1995exact,masoliver1996exact} the problem of escape of the integrated Wiener (random acceleration process\cite{theodore2014first}) process from $[0,l]$ has been studied.
For
\begin{equation}
    \ddot{x}(t)=\xi(t),
    \label{eq:mas-iw}
\end{equation}
with $\langle \xi(t) \xi(s) \rangle = D\delta(t-s)$ the closed (up to quadrature) formula for the MFPT has been derived.
In the case of $v_0=0$ the general formula simplifies to
\begin{eqnarray}
\label{eq:integratedwiener-mfpt-maso}
\mathcal{T}(x,0) & = &
\frac{4^{1/6}}{3^{1/6} \Gamma(7/3)}
\left[ \frac{2l^2}{D}  \right]^{1/3}
\left[ \frac{x}{l}  \right]^{1/6}
\left[1 - \frac{x}{l}  \right]^{1/6} \\
& \times &
\left\{
{}_2F\left(  1,-\frac{1}{3};\frac{7}{6}; \frac{x}{l}  \right)
+
{}_2F\left(  1,-\frac{1}{3};\frac{7}{6}; 1-\frac{x}{l}  \right)
\right\} \nonumber,
\end{eqnarray}
where ${}_2F\left(  a,b ; c; x  \right)$ is the Gauss hypergeometric function \cite{abramowitz1964handbook}.

The formula (\ref{eq:integratedwiener-mfpt-maso}) can be transformed to the setup used in the main text by rescaling the interval width, i.e., $l\to 2l$, and exchanging
$x \to l+x$ and $D\to 2 \sigma^2$ (because $\alpha$-stable density with $\alpha=2$ is the normal distribution with the variance equal to $2$) resulting in
\begin{eqnarray}
\label{eq:integratedwiener-mfpt}
\mathcal{T}(x,0) & = &
\frac{4^{1/6}}{3^{1/6} \Gamma(7/3)}
\left[ \frac{4l^2}{\sigma^2}  \right]^{1/3}
\left[ \frac{l+x}{2l}  \right]^{1/6}
\left[ \frac{l-x}{2l}  \right]^{1/6} \\
& \times &
\left\{
{}_2F\left(  1,-\frac{1}{3};\frac{7}{6}; \frac{l+x}{2l}  \right)
+
{}_2F\left(  1,-\frac{1}{3};\frac{7}{6}; \frac{l-x}{2l}  \right)
\right\} \nonumber.
\end{eqnarray}
Eq.~(\ref{eq:integratedwiener-mfpt}) gives the formula for the MFPT with $v_0=0$ in the dimensional units corresponding to the setup studied in the main text (escape from $[-l,l]$ under action of $\alpha$-stable noise).
Eq.~(\ref{eq:integratedwiener-mfpt}) can be transformed to dimensionless units by identifying $x/l$ with dimensionless position and dividing the whole formula by $t_0$ given by Eq.~(\ref{eq:t0}) with $m=1$, i.e., $t_0= (l/\sigma)^{2/3}$  since in \cite{masoliver1995exact,masoliver1996exact} particle mass is set to unity, see Eq.~(\ref{eq:mas-iw}).
All these operations are equivalent to setting $\sigma=1$ and $l=1$ in Eq.~(\ref{eq:integratedwiener-mfpt}).
Such a substitution is consistent with the dimensionless Langevin equation, which for the integrated $\alpha$-stable motion does not have any free parameters.


\section*{Acknowledgements}

This research was supported in part by PLGrid Infrastructure and by the National Science Center (Poland) grant 2018/31/N/ST2/00598.

%
%
\section*{Data availability}
The data that support the findings of this study are available from the corresponding author (KC) upon reasonable request.

\section*{References}

\begin{thebibliography}{61}%
\makeatletter
\providecommand \@ifxundefined [1]{%
 \@ifx{#1\undefined}
}%
\providecommand \@ifnum [1]{%
 \ifnum #1\expandafter \@firstoftwo
 \else \expandafter \@secondoftwo
 \fi
}%
\providecommand \@ifx [1]{%
 \ifx #1\expandafter \@firstoftwo
 \else \expandafter \@secondoftwo
 \fi
}%
\providecommand \natexlab [1]{#1}%
\providecommand \enquote  [1]{``#1''}%
\providecommand \bibnamefont  [1]{#1}%
\providecommand \bibfnamefont [1]{#1}%
\providecommand \citenamefont [1]{#1}%
\providecommand \href@noop [0]{\@secondoftwo}%
\providecommand \href [0]{\begingroup \@sanitize@url \@href}%
\providecommand \@href[1]{\@@startlink{#1}\@@href}%
\providecommand \@@href[1]{\endgroup#1\@@endlink}%
\providecommand \@sanitize@url [0]{\catcode `\\12\catcode `\$12\catcode
  `\&12\catcode `\#12\catcode `\^12\catcode `\_12\catcode `\%12\relax}%
\providecommand \@@startlink[1]{}%
\providecommand \@@endlink[0]{}%
\providecommand \url  [0]{\begingroup\@sanitize@url \@url }%
\providecommand \@url [1]{\endgroup\@href {#1}{\urlprefix }}%
\providecommand \urlprefix  [0]{URL }%
\providecommand \Eprint [0]{\href }%
\providecommand \doibase [0]{http://dx.doi.org/}%
\providecommand \selectlanguage [0]{\@gobble}%
\providecommand \bibinfo  [0]{\@secondoftwo}%
\providecommand \bibfield  [0]{\@secondoftwo}%
\providecommand \translation [1]{[#1]}%
\providecommand \BibitemOpen [0]{}%
\providecommand \bibitemStop [0]{}%
\providecommand \bibitemNoStop [0]{.\EOS\space}%
\providecommand \EOS [0]{\spacefactor3000\relax}%
\providecommand \BibitemShut  [1]{\csname bibitem#1\endcsname}%
\let\auto@bib@innerbib\@empty
\bibitem [{\citenamefont {Horsthemke}\ and\ \citenamefont
  {Lefever}(1984)}]{horsthemke1984}%
  \BibitemOpen
  \bibfield  {author} {\bibinfo {author} {\bibfnamefont {W.}~\bibnamefont
  {Horsthemke}}\ and\ \bibinfo {author} {\bibfnamefont {R.}~\bibnamefont
  {Lefever}},\ }\href@noop {} {\emph {\bibinfo {title} {Noise-inducted
  transitions. Theory and applications in physics, chemistry, and biology}}}\
  (\bibinfo  {publisher} {Springer Verlag},\ \bibinfo {address} {Berlin},\
  \bibinfo {year} {1984})\BibitemShut {NoStop}%
\bibitem [{\citenamefont {Gardiner}(2009)}]{gardiner2009}%
  \BibitemOpen
  \bibfield  {author} {\bibinfo {author} {\bibfnamefont {C.~W.}\ \bibnamefont
  {Gardiner}},\ }\href@noop {} {\emph {\bibinfo {title} {Handbook of stochastic
  methods for physics, chemistry and natural sciences}}}\ (\bibinfo
  {publisher} {Springer Verlag},\ \bibinfo {address} {Berlin},\ \bibinfo {year}
  {2009})\BibitemShut {NoStop}%
\bibitem [{\citenamefont {Coffey}\ and\ \citenamefont
  {Kalmykov}(2012)}]{coffey2012langevin}%
  \BibitemOpen
  \bibfield  {author} {\bibinfo {author} {\bibfnamefont {W.~T.}\ \bibnamefont
  {Coffey}}\ and\ \bibinfo {author} {\bibfnamefont {Y.~P.}\ \bibnamefont
  {Kalmykov}},\ }\href@noop {} {\emph {\bibinfo {title} {The Langevin equation:
  With applications to stochastic problems in Physics, Chemistry and Electrical
  Engineering}}}\ (\bibinfo  {publisher} {World Scientific},\ \bibinfo
  {address} {Singapore},\ \bibinfo {year} {2012})\BibitemShut {NoStop}%
\bibitem [{\citenamefont {Klafter}\ \emph {et~al.}(1987)\citenamefont
  {Klafter}, \citenamefont {Blumen},\ and\ \citenamefont
  {Shlesinger}}]{klafter1987}%
  \BibitemOpen
  \bibfield  {author} {\bibinfo {author} {\bibfnamefont {J.}~\bibnamefont
  {Klafter}}, \bibinfo {author} {\bibfnamefont {A.}~\bibnamefont {Blumen}}, \
  and\ \bibinfo {author} {\bibfnamefont {M.~F.}\ \bibnamefont {Shlesinger}},\
  }\href@noop {} {\bibfield  {journal} {\bibinfo  {journal} {Phys. Rev. A}\
  }\textbf {\bibinfo {volume} {35}},\ \bibinfo {pages} {3081} (\bibinfo {year}
  {1987})}\BibitemShut {NoStop}%
\bibitem [{\citenamefont {Klafter}\ \emph {et~al.}(1996)\citenamefont
  {Klafter}, \citenamefont {Shlesinger},\ and\ \citenamefont
  {Zumofen}}]{klafter1996}%
  \BibitemOpen
  \bibfield  {author} {\bibinfo {author} {\bibfnamefont {J.}~\bibnamefont
  {Klafter}}, \bibinfo {author} {\bibfnamefont {M.~F.}\ \bibnamefont
  {Shlesinger}}, \ and\ \bibinfo {author} {\bibfnamefont {G.}~\bibnamefont
  {Zumofen}},\ }\href@noop {} {\bibfield  {journal} {\bibinfo  {journal} {Phys.
  Today}\ }\textbf {\bibinfo {volume} {49}},\ \bibinfo {pages} {33} (\bibinfo
  {year} {1996})}\BibitemShut {NoStop}%
\bibitem [{\citenamefont {Agudov}\ and\ \citenamefont
  {Spagnolo}(2001)}]{agudov2001}%
  \BibitemOpen
  \bibfield  {author} {\bibinfo {author} {\bibfnamefont {N.~V.}\ \bibnamefont
  {Agudov}}\ and\ \bibinfo {author} {\bibfnamefont {B.}~\bibnamefont
  {Spagnolo}},\ }\href@noop {} {\bibfield  {journal} {\bibinfo  {journal}
  {Phys. Rev. E}\ }\textbf {\bibinfo {volume} {64}},\ \bibinfo {pages} {035102}
  (\bibinfo {year} {2001})}\BibitemShut {NoStop}%
\bibitem [{\citenamefont {Dubkov}\ \emph {et~al.}(2004)\citenamefont {Dubkov},
  \citenamefont {Agudov},\ and\ \citenamefont {Spagnolo}}]{dubkov2004}%
  \BibitemOpen
  \bibfield  {author} {\bibinfo {author} {\bibfnamefont {A.~A.}\ \bibnamefont
  {Dubkov}}, \bibinfo {author} {\bibfnamefont {N.~V.}\ \bibnamefont {Agudov}},
  \ and\ \bibinfo {author} {\bibfnamefont {B.}~\bibnamefont {Spagnolo}},\
  }\href@noop {} {\bibfield  {journal} {\bibinfo  {journal} {Phys. Rev. E}\
  }\textbf {\bibinfo {volume} {69}},\ \bibinfo {pages} {061103} (\bibinfo
  {year} {2004})}\BibitemShut {NoStop}%
\bibitem [{\citenamefont {Valenti}\ \emph {et~al.}(2015)\citenamefont
  {Valenti}, \citenamefont {Magazz\`u}, \citenamefont {Caldara},\ and\
  \citenamefont {Spagnolo}}]{valenti2015}%
  \BibitemOpen
  \bibfield  {author} {\bibinfo {author} {\bibfnamefont {D.}~\bibnamefont
  {Valenti}}, \bibinfo {author} {\bibfnamefont {L.}~\bibnamefont {Magazz\`u}},
  \bibinfo {author} {\bibfnamefont {P.}~\bibnamefont {Caldara}}, \ and\
  \bibinfo {author} {\bibfnamefont {B.}~\bibnamefont {Spagnolo}},\ }\href@noop
  {} {\bibfield  {journal} {\bibinfo  {journal} {Phys. Rev. B}\ }\textbf
  {\bibinfo {volume} {91}},\ \bibinfo {pages} {235412} (\bibinfo {year}
  {2015})}\BibitemShut {NoStop}%
\bibitem [{\citenamefont {Devoret}\ \emph {et~al.}(1984)\citenamefont
  {Devoret}, \citenamefont {Martinis}, \citenamefont {Esteve},\ and\
  \citenamefont {Clarke}}]{devoret1984}%
  \BibitemOpen
  \bibfield  {author} {\bibinfo {author} {\bibfnamefont {M.~H.}\ \bibnamefont
  {Devoret}}, \bibinfo {author} {\bibfnamefont {J.~M.}\ \bibnamefont
  {Martinis}}, \bibinfo {author} {\bibfnamefont {D.}~\bibnamefont {Esteve}}, \
  and\ \bibinfo {author} {\bibfnamefont {J.}~\bibnamefont {Clarke}},\ }\href
  {\doibase 10.1103/PhysRevLett.53.1260} {\bibfield  {journal} {\bibinfo
  {journal} {Phys. Rev. Lett.}\ }\textbf {\bibinfo {volume} {53}},\ \bibinfo
  {pages} {1260} (\bibinfo {year} {1984})}\BibitemShut {NoStop}%
\bibitem [{\citenamefont {Doering}\ and\ \citenamefont
  {Gadoua}(1992)}]{doering1992}%
  \BibitemOpen
  \bibfield  {author} {\bibinfo {author} {\bibfnamefont {C.~R.}\ \bibnamefont
  {Doering}}\ and\ \bibinfo {author} {\bibfnamefont {J.~C.}\ \bibnamefont
  {Gadoua}},\ }\href@noop {} {\bibfield  {journal} {\bibinfo  {journal} {Phys.
  Rev. Lett.}\ }\textbf {\bibinfo {volume} {69}},\ \bibinfo {pages} {2318}
  (\bibinfo {year} {1992})}\BibitemShut {NoStop}%
\bibitem [{\citenamefont {Evans}\ and\ \citenamefont
  {Majumdar}(2011{\natexlab{a}})}]{evans2011diffusion}%
  \BibitemOpen
  \bibfield  {author} {\bibinfo {author} {\bibfnamefont {M.~R.}\ \bibnamefont
  {Evans}}\ and\ \bibinfo {author} {\bibfnamefont {S.~N.}\ \bibnamefont
  {Majumdar}},\ }\href@noop {} {\bibfield  {journal} {\bibinfo  {journal} {Phys
  Rev. Lett.}\ }\textbf {\bibinfo {volume} {106}},\ \bibinfo {pages} {160601}
  (\bibinfo {year} {2011}{\natexlab{a}})}\BibitemShut {NoStop}%
\bibitem [{\citenamefont {Evans}\ and\ \citenamefont
  {Majumdar}(2011{\natexlab{b}})}]{evans2011diffusion-jpa}%
  \BibitemOpen
  \bibfield  {author} {\bibinfo {author} {\bibfnamefont {M.~R.}\ \bibnamefont
  {Evans}}\ and\ \bibinfo {author} {\bibfnamefont {S.~N.}\ \bibnamefont
  {Majumdar}},\ }\href@noop {} {\bibfield  {journal} {\bibinfo  {journal} {J.
  Phys. A: Math. Theor.}\ }\textbf {\bibinfo {volume} {44}},\ \bibinfo {pages}
  {435001} (\bibinfo {year} {2011}{\natexlab{b}})}\BibitemShut {NoStop}%
\bibitem [{\citenamefont {Evans}\ \emph {et~al.}(2020)\citenamefont {Evans},
  \citenamefont {Majumdar},\ and\ \citenamefont
  {Schehr}}]{evans2020stochastic}%
  \BibitemOpen
  \bibfield  {author} {\bibinfo {author} {\bibfnamefont {M.~R.}\ \bibnamefont
  {Evans}}, \bibinfo {author} {\bibfnamefont {S.~N.}\ \bibnamefont {Majumdar}},
  \ and\ \bibinfo {author} {\bibfnamefont {G.}~\bibnamefont {Schehr}},\ }\href
  {\doibase 10.1088/1751-8121/ab7cfe} {\bibfield  {journal} {\bibinfo
  {journal} {J. Phys. A: Math. Theor.}\ }\textbf {\bibinfo {volume} {53}},\
  \bibinfo {pages} {193001} (\bibinfo {year} {2020})}\BibitemShut {NoStop}%
\bibitem [{\citenamefont {Astumian}\ and\ \citenamefont
  {Moss}(1998)}]{astumian1998}%
  \BibitemOpen
  \bibfield  {author} {\bibinfo {author} {\bibfnamefont {R.~D.}\ \bibnamefont
  {Astumian}}\ and\ \bibinfo {author} {\bibfnamefont {F.}~\bibnamefont
  {Moss}},\ }\href@noop {} {\bibfield  {journal} {\bibinfo  {journal} {Chaos}\
  }\textbf {\bibinfo {volume} {8}},\ \bibinfo {pages} {533} (\bibinfo {year}
  {1998})}\BibitemShut {NoStop}%
\bibitem [{\citenamefont {Reimann}(2002)}]{reimann2002}%
  \BibitemOpen
  \bibfield  {author} {\bibinfo {author} {\bibfnamefont {P.}~\bibnamefont
  {Reimann}},\ }\href@noop {} {\bibfield  {journal} {\bibinfo  {journal} {Phys.
  Rep.}\ }\textbf {\bibinfo {volume} {361}},\ \bibinfo {pages} {57} (\bibinfo
  {year} {2002})}\BibitemShut {NoStop}%
\bibitem [{\citenamefont {Gammaitoni}\ \emph {et~al.}(2009)\citenamefont
  {Gammaitoni}, \citenamefont {H{\"a}nggi}, \citenamefont {Jung},\ and\
  \citenamefont {Marchesoni}}]{gammaitoni2009}%
  \BibitemOpen
  \bibfield  {author} {\bibinfo {author} {\bibfnamefont {L.}~\bibnamefont
  {Gammaitoni}}, \bibinfo {author} {\bibfnamefont {P.}~\bibnamefont
  {H{\"a}nggi}}, \bibinfo {author} {\bibfnamefont {P.}~\bibnamefont {Jung}}, \
  and\ \bibinfo {author} {\bibfnamefont {F.}~\bibnamefont {Marchesoni}},\
  }\href@noop {} {\bibfield  {journal} {\bibinfo  {journal} {Eur. Phys. J. B}\
  }\textbf {\bibinfo {volume} {69}},\ \bibinfo {pages} {1} (\bibinfo {year}
  {2009})}\BibitemShut {NoStop}%
\bibitem [{\citenamefont {Benichou}\ \emph {et~al.}(2005)\citenamefont
  {Benichou}, \citenamefont {Coppey}, \citenamefont {Moreau}, \citenamefont
  {Suet},\ and\ \citenamefont {Voituriez}}]{benichou2005a}%
  \BibitemOpen
  \bibfield  {author} {\bibinfo {author} {\bibfnamefont {O.}~\bibnamefont
  {Benichou}}, \bibinfo {author} {\bibfnamefont {M.}~\bibnamefont {Coppey}},
  \bibinfo {author} {\bibfnamefont {M.}~\bibnamefont {Moreau}}, \bibinfo
  {author} {\bibfnamefont {P.~H.}\ \bibnamefont {Suet}}, \ and\ \bibinfo
  {author} {\bibfnamefont {R.}~\bibnamefont {Voituriez}},\ }\href@noop {}
  {\bibfield  {journal} {\bibinfo  {journal} {EPL (Europhys. Lett.)}\ }\textbf
  {\bibinfo {volume} {70}},\ \bibinfo {pages} {42} (\bibinfo {year}
  {2005})}\BibitemShut {NoStop}%
\bibitem [{\citenamefont {Borodin}\ and\ \citenamefont
  {Salminen}(2002)}]{borodin2002}%
  \BibitemOpen
  \bibfield  {author} {\bibinfo {author} {\bibfnamefont {A.~N.}\ \bibnamefont
  {Borodin}}\ and\ \bibinfo {author} {\bibfnamefont {P.}~\bibnamefont
  {Salminen}},\ }\href@noop {} {\emph {\bibinfo {title} {Handbook of Brownian
  motion: facts and formulae}}}\ (\bibinfo  {publisher} {Birkh\"auser},\
  \bibinfo {address} {Bassel},\ \bibinfo {year} {2002})\BibitemShut {NoStop}%
\bibitem [{\citenamefont {Palyulin}\ \emph {et~al.}(2019)\citenamefont
  {Palyulin}, \citenamefont {Blackburn}, \citenamefont {Lomholt}, \citenamefont
  {Watkins}, \citenamefont {Metzler}, \citenamefont {Klages},\ and\
  \citenamefont {Chechkin}}]{palyulin2019first}%
  \BibitemOpen
  \bibfield  {author} {\bibinfo {author} {\bibfnamefont {V.~V.}\ \bibnamefont
  {Palyulin}}, \bibinfo {author} {\bibfnamefont {G.}~\bibnamefont {Blackburn}},
  \bibinfo {author} {\bibfnamefont {M.~A.}\ \bibnamefont {Lomholt}}, \bibinfo
  {author} {\bibfnamefont {N.~W.}\ \bibnamefont {Watkins}}, \bibinfo {author}
  {\bibfnamefont {R.}~\bibnamefont {Metzler}}, \bibinfo {author} {\bibfnamefont
  {R.}~\bibnamefont {Klages}}, \ and\ \bibinfo {author} {\bibfnamefont {A.~V.}\
  \bibnamefont {Chechkin}},\ }\href@noop {} {\bibfield  {journal} {\bibinfo
  {journal} {New J. Phys.}\ }\textbf {\bibinfo {volume} {21}},\ \bibinfo
  {pages} {103028} (\bibinfo {year} {2019})}\BibitemShut {NoStop}%
\bibitem [{\citenamefont {Barndorff-Nielsen}\ and\ \citenamefont
  {Shephard}(2003)}]{barndorff2003integrated}%
  \BibitemOpen
  \bibfield  {author} {\bibinfo {author} {\bibfnamefont {O.~E.}\ \bibnamefont
  {Barndorff-Nielsen}}\ and\ \bibinfo {author} {\bibfnamefont {N.}~\bibnamefont
  {Shephard}},\ }\href@noop {} {\bibfield  {journal} {\bibinfo  {journal}
  {Scand. J. Stat.}\ }\textbf {\bibinfo {volume} {30}},\ \bibinfo {pages} {277}
  (\bibinfo {year} {2003})}\BibitemShut {NoStop}%
\bibitem [{\citenamefont {Bicout}\ and\ \citenamefont
  {Burkhardt}(2000)}]{bicout2000absorption}%
  \BibitemOpen
  \bibfield  {author} {\bibinfo {author} {\bibfnamefont {D.~J.}\ \bibnamefont
  {Bicout}}\ and\ \bibinfo {author} {\bibfnamefont {T.~W.}\ \bibnamefont
  {Burkhardt}},\ }\href@noop {} {\bibfield  {journal} {\bibinfo  {journal} {J.
  Phys. A: Math. Gen.}\ }\textbf {\bibinfo {volume} {33}},\ \bibinfo {pages}
  {6835} (\bibinfo {year} {2000})}\BibitemShut {NoStop}%
\bibitem [{\citenamefont {Burkhardt}(2014)}]{theodore2014first}%
  \BibitemOpen
  \bibfield  {author} {\bibinfo {author} {\bibfnamefont {T.~W.}\ \bibnamefont
  {Burkhardt}},\ }in\ \href@noop {} {\emph {\bibinfo {booktitle} {First-Passage
  Phenomena and Their Applications}}},\ \bibinfo {editor} {edited by\ \bibinfo
  {editor} {\bibfnamefont {R.}~\bibnamefont {Metzler}}, \bibinfo {editor}
  {\bibfnamefont {G.}~\bibnamefont {Oshanin}}, \ and\ \bibinfo {editor}
  {\bibfnamefont {S.}~\bibnamefont {Redner}}}\ (\bibinfo  {publisher} {World
  Scientific Publishing},\ \bibinfo {year} {2014})\ pp.\ \bibinfo {pages}
  {21--44}\BibitemShut {NoStop}%
\bibitem [{\citenamefont {Goldman}(1971)}]{goldman1971first}%
  \BibitemOpen
  \bibfield  {author} {\bibinfo {author} {\bibfnamefont {M.}~\bibnamefont
  {Goldman}},\ }\href@noop {} {\bibfield  {journal} {\bibinfo  {journal} {Ann.
  Math. Stat.}\ }\textbf {\bibinfo {volume} {6}},\ \bibinfo {pages} {2150}
  (\bibinfo {year} {1971})}\BibitemShut {NoStop}%
\bibitem [{\citenamefont {Masoliver}\ and\ \citenamefont
  {Porr{\`a}}(1996)}]{masoliver1996exact}%
  \BibitemOpen
  \bibfield  {author} {\bibinfo {author} {\bibfnamefont {J.}~\bibnamefont
  {Masoliver}}\ and\ \bibinfo {author} {\bibfnamefont {J.~M.}\ \bibnamefont
  {Porr{\`a}}},\ }\href@noop {} {\bibfield  {journal} {\bibinfo  {journal}
  {Phys. Rev. E}\ }\textbf {\bibinfo {volume} {53}},\ \bibinfo {pages} {2243}
  (\bibinfo {year} {1996})}\BibitemShut {NoStop}%
\bibitem [{\citenamefont {Masoliver}\ and\ \citenamefont
  {Porr{\`a}}(1995)}]{masoliver1995exact}%
  \BibitemOpen
  \bibfield  {author} {\bibinfo {author} {\bibfnamefont {J.}~\bibnamefont
  {Masoliver}}\ and\ \bibinfo {author} {\bibfnamefont {J.~M.}\ \bibnamefont
  {Porr{\`a}}},\ }\href@noop {} {\bibfield  {journal} {\bibinfo  {journal}
  {Phys. Rev. Lett.}\ }\textbf {\bibinfo {volume} {75}},\ \bibinfo {pages}
  {189} (\bibinfo {year} {1995})}\BibitemShut {NoStop}%
\bibitem [{\citenamefont {Lefebvre}(1989{\natexlab{a}})}]{lefebvre1989first}%
  \BibitemOpen
  \bibfield  {author} {\bibinfo {author} {\bibfnamefont {M.}~\bibnamefont
  {Lefebvre}},\ }\href@noop {} {\bibfield  {journal} {\bibinfo  {journal} {SIAM
  J. Appl. Math.}\ }\textbf {\bibinfo {volume} {49}},\ \bibinfo {pages} {1514}
  (\bibinfo {year} {1989}{\natexlab{a}})}\BibitemShut {NoStop}%
\bibitem [{\citenamefont {Hesse}(2005)}]{hesse2005first}%
  \BibitemOpen
  \bibfield  {author} {\bibinfo {author} {\bibfnamefont {C.~H.}\ \bibnamefont
  {Hesse}},\ }\href@noop {} {\bibfield  {journal} {\bibinfo  {journal} {Int. J.
  Stoch. Anal.}\ }\textbf {\bibinfo {volume} {2005}},\ \bibinfo {pages} {237}
  (\bibinfo {year} {2005})}\BibitemShut {NoStop}%
\bibitem [{\citenamefont {Hintze}\ and\ \citenamefont
  {Pavlyukevich}(2014)}]{hintze2014small}%
  \BibitemOpen
  \bibfield  {author} {\bibinfo {author} {\bibfnamefont {R.}~\bibnamefont
  {Hintze}}\ and\ \bibinfo {author} {\bibfnamefont {I.}~\bibnamefont
  {Pavlyukevich}},\ }\href@noop {} {\bibfield  {journal} {\bibinfo  {journal}
  {Bernoulli}\ }\textbf {\bibinfo {volume} {20}},\ \bibinfo {pages} {265}
  (\bibinfo {year} {2014})}\BibitemShut {NoStop}%
\bibitem [{\citenamefont {Hesse}(1991)}]{hesse1991one}%
  \BibitemOpen
  \bibfield  {author} {\bibinfo {author} {\bibfnamefont {C.}~\bibnamefont
  {Hesse}},\ }\href@noop {} {\bibfield  {journal} {\bibinfo  {journal} {Stoch.
  Models}\ }\textbf {\bibinfo {volume} {7}},\ \bibinfo {pages} {447} (\bibinfo
  {year} {1991})}\BibitemShut {NoStop}%
\bibitem [{\citenamefont {Magdziarz}(2008)}]{magdziarz2008fractional}%
  \BibitemOpen
  \bibfield  {author} {\bibinfo {author} {\bibfnamefont {M.}~\bibnamefont
  {Magdziarz}},\ }\href@noop {} {\bibfield  {journal} {\bibinfo  {journal}
  {Physica A}\ }\textbf {\bibinfo {volume} {387}},\ \bibinfo {pages} {123}
  (\bibinfo {year} {2008})}\BibitemShut {NoStop}%
\bibitem [{\citenamefont {Srokowski}(2012)}]{srokowski2012anomalous}%
  \BibitemOpen
  \bibfield  {author} {\bibinfo {author} {\bibfnamefont {T.}~\bibnamefont
  {Srokowski}},\ }\href@noop {} {\bibfield  {journal} {\bibinfo  {journal}
  {Phys. Rev. E}\ }\textbf {\bibinfo {volume} {85}},\ \bibinfo {pages} {021118}
  (\bibinfo {year} {2012})}\BibitemShut {NoStop}%
\bibitem [{\citenamefont {Bai}\ and\ \citenamefont {Hu}(2017)}]{bai2017escape}%
  \BibitemOpen
  \bibfield  {author} {\bibinfo {author} {\bibfnamefont {Z.-W.}\ \bibnamefont
  {Bai}}\ and\ \bibinfo {author} {\bibfnamefont {M.}~\bibnamefont {Hu}},\
  }\href@noop {} {\bibfield  {journal} {\bibinfo  {journal} {Physica A}\
  }\textbf {\bibinfo {volume} {479}},\ \bibinfo {pages} {91} (\bibinfo {year}
  {2017})}\BibitemShut {NoStop}%
\bibitem [{\citenamefont {Capa{\l}a}\ and\ \citenamefont
  {Dybiec}(2020)}]{capala2020underdamped}%
  \BibitemOpen
  \bibfield  {author} {\bibinfo {author} {\bibfnamefont {K.}~\bibnamefont
  {Capa{\l}a}}\ and\ \bibinfo {author} {\bibfnamefont {B.}~\bibnamefont
  {Dybiec}},\ }\href@noop {} {\bibfield  {journal} {\bibinfo  {journal} {Phys.
  Rev. E}\ }\textbf {\bibinfo {volume} {102}},\ \bibinfo {pages} {052123}
  (\bibinfo {year} {2020})}\BibitemShut {NoStop}%
\bibitem [{\citenamefont {Dybiec}\ \emph {et~al.}(2017)\citenamefont {Dybiec},
  \citenamefont {Gudowska-Nowak}, \citenamefont {Barkai},\ and\ \citenamefont
  {Dubkov}}]{dybiec2017levy}%
  \BibitemOpen
  \bibfield  {author} {\bibinfo {author} {\bibfnamefont {B.}~\bibnamefont
  {Dybiec}}, \bibinfo {author} {\bibfnamefont {E.}~\bibnamefont
  {Gudowska-Nowak}}, \bibinfo {author} {\bibfnamefont {E.}~\bibnamefont
  {Barkai}}, \ and\ \bibinfo {author} {\bibfnamefont {A.~A.}\ \bibnamefont
  {Dubkov}},\ }\href@noop {} {\bibfield  {journal} {\bibinfo  {journal} {Phys.
  Rev. E}\ }\textbf {\bibinfo {volume} {95}},\ \bibinfo {pages} {052102}
  (\bibinfo {year} {2017})}\BibitemShut {NoStop}%
\bibitem [{\citenamefont {Redner}(2001)}]{redner2001}%
  \BibitemOpen
  \bibfield  {author} {\bibinfo {author} {\bibfnamefont {S.}~\bibnamefont
  {Redner}},\ }\href@noop {} {\emph {\bibinfo {title} {A guide to first passage
  time processes}}}\ (\bibinfo  {publisher} {Cambridge University Press},\
  \bibinfo {address} {Cambridge},\ \bibinfo {year} {2001})\BibitemShut
  {NoStop}%
\bibitem [{\citenamefont {Samorodnitsky}\ and\ \citenamefont
  {Taqqu}(1994)}]{samorodnitsky1994}%
  \BibitemOpen
  \bibfield  {author} {\bibinfo {author} {\bibfnamefont {G.}~\bibnamefont
  {Samorodnitsky}}\ and\ \bibinfo {author} {\bibfnamefont {M.~S.}\ \bibnamefont
  {Taqqu}},\ }\href@noop {} {\emph {\bibinfo {title} {Stable non-{Gaussian}
  random processes: Stochastic models with infinite variance}}}\ (\bibinfo
  {publisher} {Chapman and Hall},\ \bibinfo {address} {New York},\ \bibinfo
  {year} {1994})\BibitemShut {NoStop}%
\bibitem [{\citenamefont {Chechkin}\ \emph
  {et~al.}(2003{\natexlab{a}})\citenamefont {Chechkin}, \citenamefont
  {Metzler}, \citenamefont {Gonchar}, \citenamefont {Klafter},\ and\
  \citenamefont {Tanatarov}}]{chechkin2003b}%
  \BibitemOpen
  \bibfield  {author} {\bibinfo {author} {\bibfnamefont {A.~V.}\ \bibnamefont
  {Chechkin}}, \bibinfo {author} {\bibfnamefont {R.}~\bibnamefont {Metzler}},
  \bibinfo {author} {\bibfnamefont {V.~Y.}\ \bibnamefont {Gonchar}}, \bibinfo
  {author} {\bibfnamefont {J.}~\bibnamefont {Klafter}}, \ and\ \bibinfo
  {author} {\bibfnamefont {L.~V.}\ \bibnamefont {Tanatarov}},\ }\href@noop {}
  {\bibfield  {journal} {\bibinfo  {journal} {J. Phys. A: Math. Gen.}\ }\textbf
  {\bibinfo {volume} {36}},\ \bibinfo {pages} {L537} (\bibinfo {year}
  {2003}{\natexlab{a}})}\BibitemShut {NoStop}%
\bibitem [{\citenamefont {Koren}\ \emph
  {et~al.}(2007{\natexlab{a}})\citenamefont {Koren}, \citenamefont {Lomholt},
  \citenamefont {Chechkin}, \citenamefont {Klafter},\ and\ \citenamefont
  {Metzler}}]{koren2007}%
  \BibitemOpen
  \bibfield  {author} {\bibinfo {author} {\bibfnamefont {T.}~\bibnamefont
  {Koren}}, \bibinfo {author} {\bibfnamefont {M.~A.}\ \bibnamefont {Lomholt}},
  \bibinfo {author} {\bibfnamefont {A.~V.}\ \bibnamefont {Chechkin}}, \bibinfo
  {author} {\bibfnamefont {J.}~\bibnamefont {Klafter}}, \ and\ \bibinfo
  {author} {\bibfnamefont {R.}~\bibnamefont {Metzler}},\ }\href@noop {}
  {\bibfield  {journal} {\bibinfo  {journal} {Phys. Rev. Lett.}\ }\textbf
  {\bibinfo {volume} {99}},\ \bibinfo {pages} {160602} (\bibinfo {year}
  {2007}{\natexlab{a}})}\BibitemShut {NoStop}%
\bibitem [{\citenamefont {Koren}\ \emph
  {et~al.}(2007{\natexlab{b}})\citenamefont {Koren}, \citenamefont {Chechkin},\
  and\ \citenamefont {Klafter}}]{koren2007b}%
  \BibitemOpen
  \bibfield  {author} {\bibinfo {author} {\bibfnamefont {T.}~\bibnamefont
  {Koren}}, \bibinfo {author} {\bibfnamefont {A.~V.}\ \bibnamefont {Chechkin}},
  \ and\ \bibinfo {author} {\bibfnamefont {J.}~\bibnamefont {Klafter}},\
  }\href@noop {} {\bibfield  {journal} {\bibinfo  {journal} {Physica A}\
  }\textbf {\bibinfo {volume} {379}},\ \bibinfo {pages} {10} (\bibinfo {year}
  {2007}{\natexlab{b}})}\BibitemShut {NoStop}%
\bibitem [{\citenamefont {Sparre~Andersen}(1954)}]{sparre1954}%
  \BibitemOpen
  \bibfield  {author} {\bibinfo {author} {\bibfnamefont {E.}~\bibnamefont
  {Sparre~Andersen}},\ }\href@noop {} {\bibfield  {journal} {\bibinfo
  {journal} {Math. Scand.}\ }\textbf {\bibinfo {volume} {2}},\ \bibinfo {pages}
  {195} (\bibinfo {year} {1954})}\BibitemShut {NoStop}%
\bibitem [{\citenamefont {Sparre~Andersen}(1953)}]{sparre1953}%
  \BibitemOpen
  \bibfield  {author} {\bibinfo {author} {\bibfnamefont {E.}~\bibnamefont
  {Sparre~Andersen}},\ }\href@noop {} {\bibfield  {journal} {\bibinfo
  {journal} {Math. Scand.}\ }\textbf {\bibinfo {volume} {1}},\ \bibinfo {pages}
  {263} (\bibinfo {year} {1953})}\BibitemShut {NoStop}%
\bibitem [{\citenamefont {Dybiec}\ \emph {et~al.}(2016)\citenamefont {Dybiec},
  \citenamefont {Gudowska-Nowak},\ and\ \citenamefont
  {Chechkin}}]{dybiec2016jpa}%
  \BibitemOpen
  \bibfield  {author} {\bibinfo {author} {\bibfnamefont {B.}~\bibnamefont
  {Dybiec}}, \bibinfo {author} {\bibfnamefont {E.}~\bibnamefont
  {Gudowska-Nowak}}, \ and\ \bibinfo {author} {\bibfnamefont {A.~V.}\
  \bibnamefont {Chechkin}},\ }\href
  {http://stacks.iop.org/1751-8121/49/i=50/a=504001} {\bibfield  {journal}
  {\bibinfo  {journal} {J. Phys. A: Math. Theor.}\ }\textbf {\bibinfo {volume}
  {49}},\ \bibinfo {pages} {504001} (\bibinfo {year} {2016})}\BibitemShut
  {NoStop}%
\bibitem [{\citenamefont {Lefebvre}(1989{\natexlab{b}})}]{lefebvre1989moment}%
  \BibitemOpen
  \bibfield  {author} {\bibinfo {author} {\bibfnamefont {M.}~\bibnamefont
  {Lefebvre}},\ }\href@noop {} {\bibfield  {journal} {\bibinfo  {journal}
  {Stoch. Proc. Appl.}\ }\textbf {\bibinfo {volume} {32}},\ \bibinfo {pages}
  {281} (\bibinfo {year} {1989}{\natexlab{b}})}\BibitemShut {NoStop}%
\bibitem [{\citenamefont {Duck}\ \emph {et~al.}(1986)\citenamefont {Duck},
  \citenamefont {Marshall},\ and\ \citenamefont {Watson}}]{duck1986first}%
  \BibitemOpen
  \bibfield  {author} {\bibinfo {author} {\bibfnamefont {P.}~\bibnamefont
  {Duck}}, \bibinfo {author} {\bibfnamefont {T.}~\bibnamefont {Marshall}}, \
  and\ \bibinfo {author} {\bibfnamefont {E.}~\bibnamefont {Watson}},\
  }\href@noop {} {\bibfield  {journal} {\bibinfo  {journal} {J. Phys. A: Math.
  Gen.}\ }\textbf {\bibinfo {volume} {19}},\ \bibinfo {pages} {3545} (\bibinfo
  {year} {1986})}\BibitemShut {NoStop}%
\bibitem [{\citenamefont {Janicki}(1996)}]{janicki1996}%
  \BibitemOpen
  \bibfield  {author} {\bibinfo {author} {\bibfnamefont {A.}~\bibnamefont
  {Janicki}},\ }\href@noop {} {\emph {\bibinfo {title} {Numerical and
  statistical approximation of stochastic differential equations with
  {non-Gaussian} measures}}}\ (\bibinfo  {publisher} {Hugo Steinhaus Centre for
  Stochastic Methods},\ \bibinfo {address} {Wroc{\l}aw},\ \bibinfo {year}
  {1996})\BibitemShut {NoStop}%
\bibitem [{\citenamefont {Dubkov}\ \emph {et~al.}(2008)\citenamefont {Dubkov},
  \citenamefont {Spagnolo},\ and\ \citenamefont {Uchaikin}}]{dubkov2008}%
  \BibitemOpen
  \bibfield  {author} {\bibinfo {author} {\bibfnamefont {A.~A.}\ \bibnamefont
  {Dubkov}}, \bibinfo {author} {\bibfnamefont {B.}~\bibnamefont {Spagnolo}}, \
  and\ \bibinfo {author} {\bibfnamefont {V.~V.}\ \bibnamefont {Uchaikin}},\
  }\href@noop {} {\bibfield  {journal} {\bibinfo  {journal} {Int. J.
  Bifurcation Chaos. Appl. Sci. Eng.}\ }\textbf {\bibinfo {volume} {18}},\
  \bibinfo {pages} {2649} (\bibinfo {year} {2008})}\BibitemShut {NoStop}%
\bibitem [{\citenamefont {Getoor}(1961)}]{getoor1961}%
  \BibitemOpen
  \bibfield  {author} {\bibinfo {author} {\bibfnamefont {R.~K.}\ \bibnamefont
  {Getoor}},\ }\href@noop {} {\bibfield  {journal} {\bibinfo  {journal} {Trans.
  Am. Math. Soc.}\ }\textbf {\bibinfo {volume} {101}},\ \bibinfo {pages} {75}
  (\bibinfo {year} {1961})}\BibitemShut {NoStop}%
\bibitem [{\citenamefont {Padash}\ \emph {et~al.}(2019)\citenamefont {Padash},
  \citenamefont {Chechkin}, \citenamefont {Dybiec}, \citenamefont
  {Pavlyukevich}, \citenamefont {Shokri},\ and\ \citenamefont
  {Metzler}}]{padash2019first}%
  \BibitemOpen
  \bibfield  {author} {\bibinfo {author} {\bibfnamefont {A.}~\bibnamefont
  {Padash}}, \bibinfo {author} {\bibfnamefont {A.~V.}\ \bibnamefont
  {Chechkin}}, \bibinfo {author} {\bibfnamefont {B.}~\bibnamefont {Dybiec}},
  \bibinfo {author} {\bibfnamefont {I.}~\bibnamefont {Pavlyukevich}}, \bibinfo
  {author} {\bibfnamefont {B.}~\bibnamefont {Shokri}}, \ and\ \bibinfo {author}
  {\bibfnamefont {R.}~\bibnamefont {Metzler}},\ }\href@noop {} {\bibfield
  {journal} {\bibinfo  {journal} {J. Phys. A: Math. Theor.}\ }\textbf {\bibinfo
  {volume} {52}},\ \bibinfo {pages} {454004} (\bibinfo {year}
  {2019})}\BibitemShut {NoStop}%
\bibitem [{\citenamefont {Padash}\ \emph {et~al.}(2020)\citenamefont {Padash},
  \citenamefont {Chechkin}, \citenamefont {Dybiec}, \citenamefont {Magdziarz},
  \citenamefont {Shokri},\ and\ \citenamefont {Metzler}}]{padash2020first}%
  \BibitemOpen
  \bibfield  {author} {\bibinfo {author} {\bibfnamefont {A.}~\bibnamefont
  {Padash}}, \bibinfo {author} {\bibfnamefont {A.~V.}\ \bibnamefont
  {Chechkin}}, \bibinfo {author} {\bibfnamefont {B.}~\bibnamefont {Dybiec}},
  \bibinfo {author} {\bibfnamefont {M.}~\bibnamefont {Magdziarz}}, \bibinfo
  {author} {\bibfnamefont {B.}~\bibnamefont {Shokri}}, \ and\ \bibinfo {author}
  {\bibfnamefont {R.}~\bibnamefont {Metzler}},\ }\href {\doibase
  10.1088/1751-8121/ab9030} {\bibfield  {journal} {\bibinfo  {journal} {J.
  Phys. A: Math. Theor.}\ }\textbf {\bibinfo {volume} {53}},\ \bibinfo {pages}
  {275002} (\bibinfo {year} {2020})}\BibitemShut {NoStop}%
\bibitem [{\citenamefont {Uhlenbeck}\ and\ \citenamefont
  {Ornstein}(1930)}]{uhlenbeck1930theory}%
  \BibitemOpen
  \bibfield  {author} {\bibinfo {author} {\bibfnamefont {G.~E.}\ \bibnamefont
  {Uhlenbeck}}\ and\ \bibinfo {author} {\bibfnamefont {L.~S.}\ \bibnamefont
  {Ornstein}},\ }\href@noop {} {\bibfield  {journal} {\bibinfo  {journal}
  {Phys. Rev.}\ }\textbf {\bibinfo {volume} {36}},\ \bibinfo {pages} {823}
  (\bibinfo {year} {1930})}\BibitemShut {NoStop}%
\bibitem [{\citenamefont {Risken}(1996)}]{risken1996fokker}%
  \BibitemOpen
  \bibfield  {author} {\bibinfo {author} {\bibfnamefont {H.}~\bibnamefont
  {Risken}},\ }in\ \href {\doibase 10.1007/978-3-642-61544-3_4} {\emph
  {\bibinfo {booktitle} {The Fokker-Planck Equation: Methods of Solution and
  Applications}}}\ (\bibinfo  {publisher} {Springer-Verlag},\ \bibinfo
  {address} {Berlin},\ \bibinfo {year} {1996})\BibitemShut {NoStop}%
\bibitem [{\citenamefont {Janicki}\ and\ \citenamefont
  {Weron}(1994{\natexlab{a}})}]{janicki1994b}%
  \BibitemOpen
  \bibfield  {author} {\bibinfo {author} {\bibfnamefont {A.}~\bibnamefont
  {Janicki}}\ and\ \bibinfo {author} {\bibfnamefont {A.}~\bibnamefont
  {Weron}},\ }\href@noop {} {\bibfield  {journal} {\bibinfo  {journal} {Stat.
  Sci.}\ }\textbf {\bibinfo {volume} {9}},\ \bibinfo {pages} {109} (\bibinfo
  {year} {1994}{\natexlab{a}})}\BibitemShut {NoStop}%
\bibitem [{\citenamefont {Porr{\`a}}\ \emph {et~al.}(1994)\citenamefont
  {Porr{\`a}}, \citenamefont {Masoliver},\ and\ \citenamefont
  {Lindenberg}}]{porra1994mean}%
  \BibitemOpen
  \bibfield  {author} {\bibinfo {author} {\bibfnamefont {J.~M.}\ \bibnamefont
  {Porr{\`a}}}, \bibinfo {author} {\bibfnamefont {J.}~\bibnamefont
  {Masoliver}}, \ and\ \bibinfo {author} {\bibfnamefont {K.}~\bibnamefont
  {Lindenberg}},\ }\href@noop {} {\bibfield  {journal} {\bibinfo  {journal}
  {Phys. Rev. E}\ }\textbf {\bibinfo {volume} {50}},\ \bibinfo {pages} {1985}
  (\bibinfo {year} {1994})}\BibitemShut {NoStop}%
\bibitem [{\citenamefont {Chechkin}\ \emph {et~al.}(2002)\citenamefont
  {Chechkin}, \citenamefont {Klafter}, \citenamefont {Gonchar}, \citenamefont
  {Metzler},\ and\ \citenamefont {Tanatarov}}]{chechkin2002}%
  \BibitemOpen
  \bibfield  {author} {\bibinfo {author} {\bibfnamefont {A.~V.}\ \bibnamefont
  {Chechkin}}, \bibinfo {author} {\bibfnamefont {J.}~\bibnamefont {Klafter}},
  \bibinfo {author} {\bibfnamefont {V.~Y.}\ \bibnamefont {Gonchar}}, \bibinfo
  {author} {\bibfnamefont {R.}~\bibnamefont {Metzler}}, \ and\ \bibinfo
  {author} {\bibfnamefont {L.~V.}\ \bibnamefont {Tanatarov}},\ }\href@noop {}
  {\bibfield  {journal} {\bibinfo  {journal} {Chem. Phys.}\ }\textbf {\bibinfo
  {volume} {284}},\ \bibinfo {pages} {233} (\bibinfo {year}
  {2002})}\BibitemShut {NoStop}%
\bibitem [{\citenamefont {Chechkin}\ \emph
  {et~al.}(2003{\natexlab{b}})\citenamefont {Chechkin}, \citenamefont
  {Klafter}, \citenamefont {Gonchar}, \citenamefont {Metzler},\ and\
  \citenamefont {Tanatarov}}]{chechkin2003}%
  \BibitemOpen
  \bibfield  {author} {\bibinfo {author} {\bibfnamefont {A.~V.}\ \bibnamefont
  {Chechkin}}, \bibinfo {author} {\bibfnamefont {J.}~\bibnamefont {Klafter}},
  \bibinfo {author} {\bibfnamefont {V.~Y.}\ \bibnamefont {Gonchar}}, \bibinfo
  {author} {\bibfnamefont {R.}~\bibnamefont {Metzler}}, \ and\ \bibinfo
  {author} {\bibfnamefont {L.~V.}\ \bibnamefont {Tanatarov}},\ }\href@noop {}
  {\bibfield  {journal} {\bibinfo  {journal} {Phys. Rev. E}\ }\textbf {\bibinfo
  {volume} {67}},\ \bibinfo {pages} {010102(R)} (\bibinfo {year}
  {2003}{\natexlab{b}})}\BibitemShut {NoStop}%
\bibitem [{\citenamefont {Dybiec}\ \emph {et~al.}(2007)\citenamefont {Dybiec},
  \citenamefont {Gudowska-Nowak},\ and\ \citenamefont {Sokolov}}]{dybiec2007d}%
  \BibitemOpen
  \bibfield  {author} {\bibinfo {author} {\bibfnamefont {B.}~\bibnamefont
  {Dybiec}}, \bibinfo {author} {\bibfnamefont {E.}~\bibnamefont
  {Gudowska-Nowak}}, \ and\ \bibinfo {author} {\bibfnamefont {I.~M.}\
  \bibnamefont {Sokolov}},\ }\href@noop {} {\bibfield  {journal} {\bibinfo
  {journal} {Phys. Rev. E}\ }\textbf {\bibinfo {volume} {76}},\ \bibinfo
  {pages} {041122} (\bibinfo {year} {2007})}\BibitemShut {NoStop}%
\bibitem [{\citenamefont {van Kampen}(1981)}]{vankampen1981}%
  \BibitemOpen
  \bibfield  {author} {\bibinfo {author} {\bibfnamefont {N.~G.}\ \bibnamefont
  {van Kampen}},\ }\href@noop {} {\emph {\bibinfo {title} {Stochastic processes
  in physics and chemistry}}}\ (\bibinfo  {publisher} {North--Holland},\
  \bibinfo {address} {Amsterdam},\ \bibinfo {year} {1981})\BibitemShut
  {NoStop}%
\bibitem [{\citenamefont {Garbaczewski}\ and\ \citenamefont
  {Olkiewicz}(2000)}]{garbaczewski2000}%
  \BibitemOpen
  \bibfield  {author} {\bibinfo {author} {\bibfnamefont {P.}~\bibnamefont
  {Garbaczewski}}\ and\ \bibinfo {author} {\bibfnamefont {R.}~\bibnamefont
  {Olkiewicz}},\ }\href@noop {} {\bibfield  {journal} {\bibinfo  {journal} {J.
  Math. Phys.}\ }\textbf {\bibinfo {volume} {41}},\ \bibinfo {pages} {6843}
  (\bibinfo {year} {2000})}\BibitemShut {NoStop}%
\bibitem [{\citenamefont {Eliazar}\ and\ \citenamefont
  {Klafter}(2005)}]{eliazar2005levy}%
  \BibitemOpen
  \bibfield  {author} {\bibinfo {author} {\bibfnamefont {I.}~\bibnamefont
  {Eliazar}}\ and\ \bibinfo {author} {\bibfnamefont {J.}~\bibnamefont
  {Klafter}},\ }\href@noop {} {\bibfield  {journal} {\bibinfo  {journal} {J.
  Stat. Phys.}\ }\textbf {\bibinfo {volume} {119}},\ \bibinfo {pages} {165}
  (\bibinfo {year} {2005})}\BibitemShut {NoStop}%
\bibitem [{\citenamefont {Janicki}\ and\ \citenamefont
  {Weron}(1994{\natexlab{b}})}]{janicki1994}%
  \BibitemOpen
  \bibfield  {author} {\bibinfo {author} {\bibfnamefont {A.}~\bibnamefont
  {Janicki}}\ and\ \bibinfo {author} {\bibfnamefont {A.}~\bibnamefont
  {Weron}},\ }\href@noop {} {\emph {\bibinfo {title} {Simulation and chaotic
  behavior of $\alpha$-stable stochastic processes}}}\ (\bibinfo  {publisher}
  {Marcel Dekker},\ \bibinfo {address} {New York},\ \bibinfo {year}
  {1994})\BibitemShut {NoStop}%
\bibitem [{\citenamefont {Abramowitz}\ and\ \citenamefont
  {Stegun}(1964)}]{abramowitz1964handbook}%
  \BibitemOpen
  \bibfield  {author} {\bibinfo {author} {\bibfnamefont {M.}~\bibnamefont
  {Abramowitz}}\ and\ \bibinfo {author} {\bibfnamefont {I.~A.}\ \bibnamefont
  {Stegun}},\ }\href@noop {} {\emph {\bibinfo {title} {Handbook of mathematical
  functions with formulas, graphs, and mathematical tables}}},\ Vol.~\bibinfo
  {volume} {55}\ (\bibinfo  {publisher} {Dover Publications},\ \bibinfo
  {address} {New York},\ \bibinfo {year} {1964})\BibitemShut {NoStop}%
\end{thebibliography}

\def\url#1{}

\end{document}